\documentclass{WileyMSP-template}

\usepackage{amsmath}
\usepackage{graphicx}

\usepackage{times}
\usepackage{dsfont}
\usepackage[pdftex]{xcolor}

\usepackage{bm}
\DeclareBoldMathCommand{\Cov}{\mathsf{C}}
\DeclareBoldMathCommand{\S}{\mathsf{S}} 

\newcommand{\Chi}{K}

\newcommand{\oMega}{\omega}

\begin{document}

\pagestyle{fancy}
\addtolength{\textheight}{-12ex}
\rhead{\includegraphics[width=2.5cm]{vch-logo.png}}

\title{Heat transfer and entanglement --
non-equilibrium correlation spectra of two quantum oscillators}

\maketitle


\author{Carsten Henkel*}


\dedication{}

\begin{affiliations}
Prof. Dr. Carsten Henkel\\
University of Potsdam, Institute of Physics and Astronomy,\\
Karl-Liebknecht-Str. 24/25, 14476 Potsdam, Germany\\
Email Address: henkel@uni-potsdam.de
\end{affiliations}



\begin{abstract}
The non-equilibrium state of two oscillators with a mutual interaction and coupled
to separate heat baths is discussed. Bosonic baths are considered, and an exact
spectral representation for the elements of the covariance matrix is provided
analytically. A wide class of spectral densities for the relevant bath modes 
is allowed for. The validity of the fluctuation-dissipation theorem is established 
for global equilibrium (both baths at the same temperature) in the stationary state. 
Spectral measures of entanglement are suggested by comparing to the equilibrium spectrum 
of zero-point fluctuations.
No rotating-wave approximation is applied, and anomalous heat transport from 
cold to hot bath, as reported in earlier work, is demonstrated not to occur.
\end{abstract}


\addtolength{\parindent}{2em}
\addtolength{\parskip}{1ex}

\section{Introduction}

The advent of quantum technology has triggered a re-analysis of thermodynamic concepts,
producing many examples of deviations and anomalies. The intimate connection between
a system state and information about it that is characteristic for quantum physics, has
provided the basis for engines that transform information into work 
\cite{Kish2012}. The thermodynamic viewpoint has also revigorated to interest in open quantum
systems, in contact with a bath and being monitored, possibly continuously.
The dynamics of open quantum systems was traditionally formulated with the help of
master equations \cite{BreuerPetruccioneBook} or influence functionals \cite{WeissBook}. 
We are witnessing a renewed discussion
about the details how these equations of motion should be set up, in order to respect
basic requirements of both thermodynamics and quantum mechanics \cite{Ford2006c, Stockburger2016, Shavit2019}.
A system coupled to
two different heat baths provides a paradigmatic example: it can be used to study heat
transport, but also leads to interesting non-equilibrium states. In the presence of
strong coupling to a bath, the separation, e.g. in terms of energy, between system and
bath gets blurred. It is no longer obvious that a given dynamical
model (master equation, Langevin equation) should be judged by its ability to generate
the stationary system state that is expected from the Boltzmann-Gibbs canonical ensemble
for the isolated system. 
After all, energy levels are being broadened and shifted by the contact with the bath,
so that a careful procedure of removing the contact to the bath should be applied.
The typical assumptions for baths
used in deriving master equations actually complicate the situation: a continuous set of 
oscillators is by itself unable to thermalise, for example, and a bath in a thermal equilibrium 
state has to be specified by its initial conditions. In this context, master equations
cannot, in general, be in Markov form because of memory kernels whose characteristic
time constants are set by the spectral density of the bath modes in so far they couple
to the system.

The background for the present study is provided by two contributions. 
Dorofeyev \cite{Dorofeyev2013a} has given for a system of two coupled oscillators coupled to
bosonic baths a spectral representation of the interaction energy. This spectrum provides
a way to visualise the heat flow across the two normal modes of the coupled oscillators.
Levy and Kosloff \cite{Levy2014} proposed a very similar model and showed that the parameters
of the two-oscillator system can be such that heat is flowing from cold to hot, in stark
violation of the second law. This has been taken as indication that one should abandon 
the concept of ``local coupling'' to a bath (each oscillator couples to its ``own bath'', 
not correlated with the bath coupled to the other one), and replace it by a ``global''
approach where the two oscillators are considered as a one system first. In the aftermath
of this paper, the issue of ``local'' vs. ``global'' couplings has been tested on various
systems \cite{Stockburger2016, Rivas2010, Martinez2013, Joshi2014, Hofer2017, OnamGonzalez2017, Farina2020}.

A related question arises in the interplay with dissipation on the one hand, as described by
a master equation in Lindblad form, and additional interactions on the other,
that may be added to the
Hamiltonian part of the master equation \cite{Shavit2019}. 
The problem is also rooted in memory effects \cite{Haake1983,Haake1985,Suarez1992},
in particular when the bath has a structured spectrum and the precise values of Bohr
frequencies are relevant, as these are shifted either by the bath or by additional 
interactions. Complications then arise with respect to the order in which the secular 
and the Markov approximations are applied \cite{Shavit2019}.
It has been argued that the Markov approximation behind 
some master equations is to blame. The advantage of the quantum Langevin formalism 
used here is that it is naturally non-Markovian, as
soon as the memory kernels are of nonzero range -- which is, of course, the
generic case. 

We study here an extremely simple situation that permits calculations with no approximations
and avoids a few of the delicate issues mentioned before. We revisit the two coupled oscillators
studied by Dorofeyev and Levy and Kosloff, but use a slightly different coupling that
preserves positivity of the Hamiltonian and is not restricted to near-degenerate resonance
frequencies,
\begin{equation}
V = \frac{\lambda}{2} ( x - y )^2
\label{eq:Potsdam-coupling}
\end{equation}
where $x$, $y$ are the oscillator coordinates. 
The Heisenberg equations of motion are worked out and take the form of exact
quantum Langevin equations \cite{Talkner1981}. For this non-Markovian system out of thermal
equilibrium, we compute the covariance matrix of the 
position and momentum canonical coordinates (also known as continuous variables in quantum
information). The formalism easily affords us with spectral representations of two-time
correlations in the long-time limit. The heat current flowing
through the system can be found by an
energy balance argument, very similar to Dorofeyev's analysis. We find that it is always
directed from the hot to the cold bath, under very general assumptions. (The baths may
have any spectral density, the damping kernels may have any memory, the bath-induced friction
may be strong or weak compared to the interaction between oscillators.) 
The anomalous heat current of Levy \& Kosloff~\cite{Levy2014} 
thus appears to be an artefact of the rotating-wave approximation
made in the interaction
\begin{equation}
V = \epsilon \left( a^\dag b + a b^\dag \right) = \lambda \, x y + \lambda' \, p_x p_y
\label{eq:rwa-coupling}
\end{equation}
where $a$, $b$ are the bosonic lowering operators constructed from $x$ and $y$ and their
conjugate momenta and $\lambda, \lambda'\propto  \epsilon$.  As an additional check of the consistency of
the results, we consider equilibrium conditions with both baths having the same temperature,
other parameters remaining arbitrary. The two-oscillator system then reaches a state 
where its two-time correlation functions satisfy the fluctuation--dissipation theorem.  The
quantum Langevin model thus relaxes the system (whatever the initial
conditions) towards thermal equilibrium at long times. 
The correlation functions in this equilibrium state \emph{differ} 
from those obtained for a canonical density operator $\exp( - H_{12} / T ) / Z$
because of bath-induced friction, a well-known 
feature of the fluctuation--dissipation theorem for strongly damped systems 
\cite{Talkner1981,Grabert1984,Goyal2019}.
The Boltzmann-Gibbs state might be reached
by disconnecting the baths (weak-coupling regime), although even that requires some 
``decoupling work'' to be done \cite{Ford2006c}.

In a second step, we consider entanglement measures computed from the covariance matrix 
at large times. We point out that the criterion of a positive partially transposed covariance
matrix may be used to construct a pair of canonical coordinates whose variances apparently
drop below the Heisenberg limit, in close analogy to the large-distance correlations
analysed by Einstein, Podolsky, and Rosen \cite{Einstein1935}. This construction can be
carried over to the spectral domain, using techniques similar to those introduced
by Ekstein and Rostoker \cite{Ekstein1955} and familiar in filter theory
(in the sense of Wagner and Campbell for signal processing \cite{Meyer2017book,Zverev1969}).
We can thus formulate a protocol that identifies the frequency band where the non-classical
correlations (entanglement) between the two oscillators can be detected with the best
margin.

\section{Model}

\subsection{Hamiltonian}

For completeness, we spell out in this section the Hamiltonian of the system. Readers
familiar with the Langevin equation may jump directly to Sec.\,\ref{s:Langevin}.

The two oscillators are described by the Hamiltonian \cite{Dorofeyev2013a,Hsiang2015}
\begin{equation}
H_{12} = 
  \frac{ p_x^2 }{ 2 m_1 } + \frac{ k_1 }{ 2 } x^2
+ \frac{ p_y^2 }{ 2 m_2 } + \frac{ k_2 }{ 2 } y^2
+ \frac12\lambda (x - y)^2
\label{eq:def-H-12}
\end{equation}
with the bilinear interaction $V_m(x - y)$ of Eq.\,(\ref{eq:Potsdam-coupling}).
The obvious notation is based on mechanical oscillators with displacements $x$, $y$,
but the model can be re-framed easily to electric systems like an LC circuit
\begin{equation*}
\frac{ \Phi^2 }{ 2 L } + \frac{ C }{ 2 } Q^2
\end{equation*}
where $Q$ is the charge on a capacity with capacitance $C > 0$,
$\Phi = L \dot Q$ the magnetic flux, and $L > 0$ the circuit 
(self-)\nolinebreak[0]inductance. In this language, the coupling $V_m$ in 
Eq.\,(\ref{eq:def-H-12}) would be called capacitive. The canonical commutator
$[\Phi, Q] = {\rm i}\hbar$ then yields the magnetic flux quantum $\hbar/e$.

Each oscillator is coupled to a bosonic bath, i.e. a collection of oscillators
\begin{equation}
H_{B1} = \sum_{j \in {\rm B}1} \bigg(
\frac{ p_j^2 }{ 2 m_j }
+ \frac{ k_j }{ 2 } ( q_j - c_j x )^2 \bigg)
\label{eq:H-bath}
\end{equation}
where $c_j$ is a (dimensionless) coupling constant and B1 represents the bath modes.
For one oscillator, this would correspond to the Ullersma \cite{Ullersma1966}
or Caldeira--Leggett \cite{Caldeira1983b} model. The dynamics becomes irreversible if
we go to the continuum limit where the spectral density
\begin{equation}
\rho_1( \oMega ) = \frac{ \pi }{ 2 } \sum_{j \in {\rm B}1} k_j c_j^2 \,\delta( \oMega - \oMega_j )
\label{eq:def-bath-rho}
\end{equation}
with $\oMega_j^2 = k_j / m_j$
becomes a smooth function. 
(This definition of the spectral density for the oscillator-bath coupling 
follows the convention of Ref.\,\cite{Dorofeyev2013a}.)
We assume that the spring constants $k_j$ and couplings $c_j$
are such that
$\rho_1( \oMega )$ smoothly decays to zero in the UV. (For this reason, 
$\rho_1( \oMega )$ differs from the bath density of states by more than just 
some power of the frequency.)
We fix an initial time $t = 0$ where the bath coordinates have equilibrated with 
the (`clamped') positions $x(0)$ and $y(0)$ of the oscillators. This leads to the
correlations
\begin{equation}
\langle p_j(0) \rangle_T = 0 =
\langle q_j(0) - x(0) \rangle_T
\label{eq:mean-values}
\end{equation}
where $\langle \ldots \rangle_T$ denotes the average at the bath temperature. Defining
the symmetrised covariances for any two bath operators
\begin{equation}
\langle A, B \rangle_T := \tfrac12 \langle A B + B A \rangle_T -
\langle A \rangle_T \langle B\rangle_T
\label{eq:def-covariance}
\end{equation}
their values in thermal equilibrium take the form
\begin{eqnarray}
\frac{k_j}{2} \langle q_j, q_j \rangle_T &=& 
\frac{k_j}{2} \langle [q_j - x(0)]^2 \rangle_T =
\tfrac12 \vartheta(\oMega_j) 
\,,
\nonumber\\
\frac{ \langle p_j, p_j \rangle_T }{ 2 m_j }  &=& 
\tfrac12 \vartheta(\oMega_j) 
\label{eq:bath-covariances}
\end{eqnarray}
The mixed symmetrised correlation $\langle q_j, p_k \rangle_T$ vanishes,
and different normal modes $j \ne k$ are not correlated:
$\langle q_j, q_k \rangle_T = 0 = \langle p_j, p_k \rangle_T$. 
Here, the effective temperature is (we set the Boltzmann constant $k_B = 1$)
\begin{equation}
\vartheta( \oMega ) 
= \frac{ \hbar\oMega }{ 2  } 
  \coth \frac{ \hbar\oMega }{ 2  T }
= \hbar\oMega 
  \left[ \bar{n}( \oMega ) + \tfrac{1}{2} \right]
\label{eq:def-theta-coth}
\end{equation}
where $\bar{n}( \oMega )$ is the Bose-Einstein distribution and $\tfrac{1}{2}
\hbar \oMega$ the zero-point energy.
In the high-temperature (or low-frequency) limit, $\vartheta( \oMega )$ reaches the classical 
(equipartition) value $T$.
By averaging
the commutator $[ q_j, p_j ] = q_j p_j - p_j q_j = {\rm i} \hbar$,
one gets the non-symmetric correlations
\begin{equation}
\left\langle \left(q_j - x(0) \right) p_j \right\rangle_T = 
- \left\langle p_j \left(q_j - x(0)\right) \right\rangle_T = 
\frac{ {\rm i} \hbar }{ 2 }
 \label{eq:}
\end{equation}
Similar expressions describe the bath attached to the other oscillator;
its spectral density will be denoted $\rho_2(\oMega)$. 
If the baths have different temperatures $T_2 \ne T_1$, this allows for a nonzero
heat current. The key assumption of this model is that the initial conditions for
the dynamical variables of bath~1 and bath~2 show no cross-correlations (`local bath').

\subsection{Langevin equations}
\label{s:Langevin}

The elimination of the bath coordinates outlined in 
Appendix\,\ref{a:calculate-Langevin-eqns} leads to the pair of Langevin 
equations
\begin{eqnarray}
\dot p_x + k_1' x &=& \lambda y - \mu_1 \!* \dot x + F_1(t)
\label{eq:Langevin-px}
\\
\dot p_y + k_2' y &=& \lambda x - \mu_2 \!* \dot y + F_2(t)
\label{eq:Langevin-py}
\end{eqnarray}
Here, we recognise with the spring constant $\lambda$ the force exerted mutually 
by the oscillators. It also modifies the oscillators' spring constants according to 
$k_i' = k_i + \lambda$ ($i = 1,2$). 
The friction force $\mu_1 \!* \dot x$ is the convolution
\begin{equation}
(\mu_1 \!* \dot x)(t) = \int\!{\rm d}\tau \, \mu_1( \tau ) \dot x( t - \tau )
\label{eq:}
\end{equation}
(analogously for $\mu_2 \!* \dot y$),
and its kernel $\mu_1( \tau )$ given by
\begin{equation}
\mu_1( \tau )
=
\Theta(\tau)
\int_0^\infty\!\frac{ {\rm d}\oMega }{ \pi/2 }
\rho_1( \oMega ) \cos\oMega\tau
\label{eq:spectrum-of-mu}
\end{equation}
The friction force is causal so that only past values $t - \tau \le t$ of
the velocity are contributing. The familiar Ohmic (memoryless) case
corresponds to $\rho_1( \oMega ) = \Gamma_1$ with the usual friction
coefficient $\Gamma_1$, since in this limit, $\mu_1 \!* \dot x = \Gamma_1 \dot x$.
(The step function $\Theta(\tau)$ cuts off one half of the $\delta(\tau)$.)
Defining the Fourier transform by 
\begin{equation}
\mu_i(\oMega) = \int\!{\rm d}\tau\,
{\rm e}^{ {\rm i} \oMega \tau }
\mu_i(\tau)
\label{eq:def-Fourier}
\end{equation}
we note the relation 
\begin{equation}
\mathop{\rm Re}\mu_i(\oMega) = \rho_i( \oMega ).
\label{eq:re-mu-and-DOS}
\end{equation}
This strictly holds only for $\oMega > 0$, but since $\mu_i(\oMega)$ describes a
response between real quantities, $\mathop{\rm Re}\mu_i(\oMega)$ is an even function 
along the real axis. It is thus convenient to consider $\rho_i(\oMega)$ even 
in  $\oMega$. Furthermore, $\mu_i(\tau)$ being a causal kernel (vanishing for $\tau < 0$), 
its Fourier transform is analytic
in the upper half of the complex frequency plane (Titchmarsh theorem, Kramers--Kronig
relations).

The Langevin force $F_1(t)$ vanishes on average and its autocorrelation function 
has a similar representation in terms of the bath spectrum
\begin{equation}
\langle F_1(t), F_1(t') \rangle_{1} =
\frac{ \hbar }{ 2 }
\int_0^\infty\!\frac{{\rm d}\oMega }{ \pi/2 }
\rho_1( \oMega ) 
\oMega \coth\frac{ \hbar \oMega }{ 2 T_1 }
\cos\oMega (t' - t)
\label{eq:Langevin-correlations}
\end{equation}
Here, $\langle \ldots \rangle_1$ denotes the average with respect to the temperature
of bath~1. The effective temperature $\vartheta(\oMega)$ defined in 
Eq.\,(\ref{eq:def-theta-coth}) is thus evaluated with $T_1$.
The Fourier transformed Langevin force
will be used in Sec.\,\ref{s:correlation-spectra}
to generate averages and correlation spectra.
This is somewhat symbolic since $F_1(\oMega)$ does
not exist in the conventional sense. By taking the double Fourier
transform of the force-force correlation function, we get, however
\begin{align}
\langle F^\dagger_1(\oMega), F_1(\oMega') \rangle_1 &=
\int\!{\rm d}t\,{\rm d}t'\,
{\rm e}^{ {\rm i} (\oMega' t' - \oMega t) }
\langle F_1(t), F_1(t') \rangle
\nonumber\\
&=
2\pi \delta( \oMega' - \oMega )
\int\!{\rm d}\tau\, 
{\rm e}^{ {\rm i} \oMega \tau }
\langle F_1(t), F_1(t + \tau) \rangle_1
\nonumber\\
&=
\pi \delta( \oMega' - \oMega )
S_{F1}( \oMega )
\label{eq:result-corr-FF-Fourier}
\end{align}
since the bath correlations are stationary and depend only on the time 
difference. (At this point, we consider $t$ and $t'$ to be in the late future
of the initial time.) The $\tau$-integral over the correlation function exists and 
defines the spectral density $S_{F1}(\oMega)$ of the Langevin force
(Wiener-Khintchine theorem~\cite{MandelWolfBook}). Since this
correlation function is real and even,
the same is true for the spectral density, and we get
by comparison to Eq.\,(\ref{eq:Langevin-correlations})
\begin{equation}
S_{F1}( \oMega ) = 4 \rho_1( \oMega ) \vartheta_1( \oMega )
\label{eq:FD1-Langevin-force}
\end{equation}
which is the fluctuation--dissipation relation for the Langevin force \cite{WeissBook}.
The prefactors in Eq.\,(\ref{eq:result-corr-FF-Fourier}, 
\ref{eq:FD1-Langevin-force})
arise from the convention that $S_F$
represents the force autocorrelation by an integral over positive frequencies
only [see Eq.\,(\ref{eq:Langevin-correlations})].

Analoguous formulas apply for the Langevin force $F_2(t)$. The assumption that each oscillator
couples to its local bath implies that there are no cross-correlations 
$\langle F_1(t), F_2(t') \rangle = 0$.

\subsection{Remarks}

The Langevin forces $F_1(t)$, $F_2(t)$ are defined in such a way that they depend on 
the initial conditions of the baths whose joint state is assumed to factorise. As the
system evolves, however, correlations arise among the two baths due to the coupling
$\lambda$ between the oscillators. Similarly, the friction forces arise because
the attached oscillators `polarise' their baths 
[the inhomogeneous term in Eq.\,(\ref{eq:solution-bath-mode})]. These features are used
in input-output theory \cite{GardinerZollerBook} to analyse the information
about the system made available in the bath variables (rather than ignoring these
as unobservable).



We already mentioned the simple Ohmic case of memoryless friction, but this is
does not necessarily imply a white noise spectrum for the Langevin force. Indeed,
only in the high-temperature (low-frequency, or classical) limit does the force
spectrum have the same frequency scaling as the friction kernel, as is also apparent from
the fluctuation--dissipation relation~(\ref{eq:FD1-Langevin-force}).
In particular in the quantum limit $\coth(\hbar \omega/2 T) \to \mathop{\rm sign}(\omega)$, 
the quantum
noise spectrum is potentially wider. The coupling constants $c_j$ then play
an essential role in determining the bath correlation time -- the latter is notably set
by the width of the relevant mode spectrum that actually couples to the system.

Dorofeyev~\cite{Dorofeyev2013a} and
Ghesqui\`ere et al.\,\cite{Ghesquiere2012}
choose the interaction
in the bilinear form $- \lambda \, x y$ and have to deal with 
instabilities at couplings $\lambda^2 \ge k_1 k_2$ where the potential surface
becomes a saddle and the system can escape to infinity along the directions
$x = y$ in the $xy$-plane. This is avoided with the positive definite 
interaction~(\ref{eq:Potsdam-coupling}), the price to pay being the shift
$k_i' = k_i + \lambda$ in the oscillator spring constants. The stability
condition $\lambda^2 < k'_1 k'_2$ then holds for any positive spring
constant $\lambda$ (and even for $0 \ge \lambda > -k_1 k_2/(k_1 + k_2)$).

\section{Correlation spectra}
\label{s:correlation-spectra}

\subsection{Representation in Fourier space}
\label{s:Fourier}

In this section, we solve the Langevin equations~(\ref{eq:Langevin-px},
\ref{eq:Langevin-py}) by a Fourier transformation and compute expectation
values. Since the friction kernels are in convolution form, we find
\begin{equation}
\begin{pmatrix}
\Chi_1(\oMega) & -\lambda
\\
-\lambda & \Chi_2(\oMega)
\end{pmatrix}
\begin{pmatrix}
x(\oMega) \\
y(\oMega)
\end{pmatrix} = 
\begin{pmatrix}
F_1(\oMega) \\
F_2(\oMega)
\end{pmatrix}
\label{eq:Fourier-matrix}
\end{equation}
Here, the diagonal terms (we may call them inverse susceptibilities) are
\begin{equation}
\Chi_1(\oMega) = 
- m_1 \oMega^2 - {\rm i} \oMega \mu_1(\oMega) + k_1'
\label{eq:def-chi}
\end{equation}
and $\mu_1(\oMega)$ is the Fourier transform of the friction 
kernel~(\ref{eq:spectrum-of-mu}).

The solution of the linear system~(\ref{eq:Fourier-matrix}) is immediate
\begin{equation}
\begin{pmatrix}
x(\oMega) \\
y(\oMega)
\end{pmatrix} = 
\frac{ 1 }{ D( \oMega ) }
\begin{pmatrix}
\Chi_2(\oMega) & \lambda
\\
\lambda & \Chi_1(\oMega)
\end{pmatrix}
\begin{pmatrix}
F_1(\oMega) \\
F_2(\oMega)
\end{pmatrix}
\label{eq:linear-response-matrix}
\end{equation}
and involves the determinant
\begin{equation}
D( \oMega ) = \Chi_1( \oMega ) \Chi_2( \oMega ) - \lambda^2
\label{eq:}
\end{equation}
Its zeros determine the eigenmodes of the coupled oscillator system.
In the simple case that damping is negligible, one gets
\begin{equation}
\oMega^2_{\pm} \approx \frac{ \oMega^{2}_1 + \oMega^{2}_2 }{ 2 }
\pm
\frac{1}{2} \left[
(\oMega^{2}_1 - \oMega^{2}_2)^2
 + 4 g^4 
\right]^{1/2}
\label{eq:eigenfrequencies}
\end{equation}
where $\oMega^{2}_i = k_i' / m_i$ ($i = 1,2$) are the eigenfrequencies.
They are shifted and pushed apart by the coupling $g^2 = \lambda / \sqrt{ m_1 m_2 }$.
Another simple case where the fourth-order characteristic polynomial simplifies
is the near-resonant one $\omega \sim \omega_{1,2}$ where the rotating-wave approximation 
can be applied.
The susceptibilities are then linearised,
$K_1(\omega) \approx 2 m_1 \omega_1 ( \omega_1 - {\rm i} \gamma_1/2 - \omega )$
with $\gamma_1 = \mu_1(\omega_1) / m_1$. The secular frequencies are found as
\begin{equation}
\omega_{\pm} \approx \frac{ \Omega_1 + \Omega_2}{2} \pm 
\frac{1}{2}
\left[ (\Omega_1 - \Omega_2)^2 + g_{r}^2 \right]^{1/2}
\label{eq:RWA-eigenfrequencies}
\end{equation}
with the complex bare resonances $\Omega_1 = \omega_1 - {\rm i} \gamma_1/2$
and the effective coupling $g_{r} = \lambda / \sqrt{m_1 m_2 \omega_1 \omega_2}$.
In particular in optical spectroscopy, one speaks of a strongly coupled system
when $g_{r}$ exceeds the linewidths $\mathop{\rm Re}(\gamma_1 + \gamma_2)$ 
so that the complex
resonances $\omega_{\pm}$ are well-resolved. The near-resonant approximation
neglects the spectral structure of the bath (and the concomitant memory effects)
by evaluating $\mu_1(\omega_1)$ at
the resonance frequency. A more complete evaluation would lead to shifted and
additional poles.

\begin{figure}
\includegraphics*[width=0.45\columnwidth]{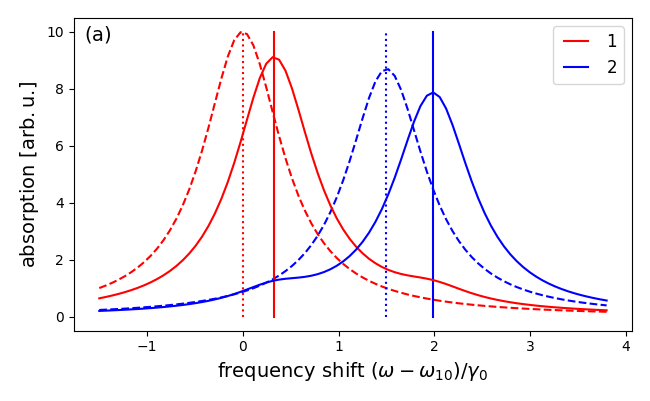}%
\hspace*{02mm}%
\includegraphics*[width=0.45\columnwidth]{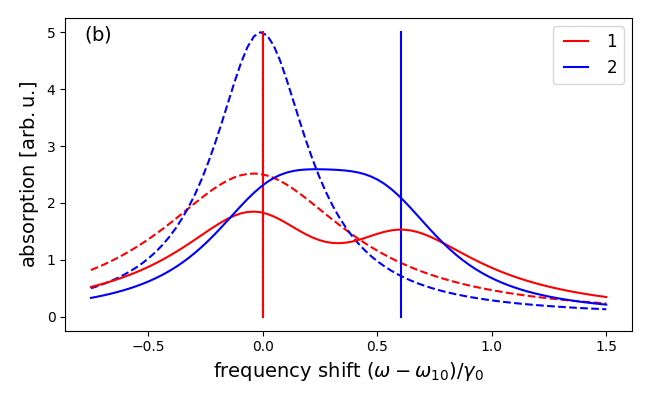}
\\
\includegraphics*[width=0.45\columnwidth]{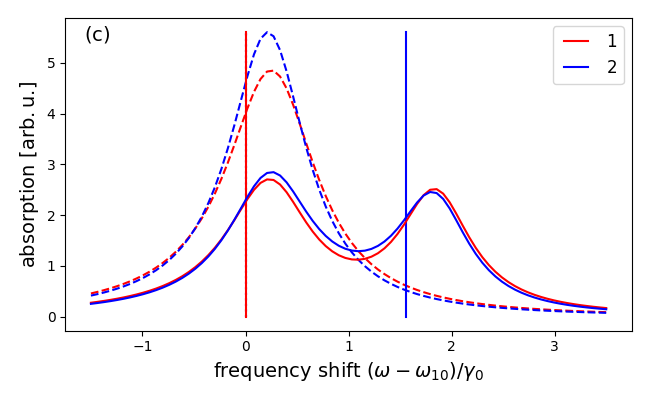}%
\hspace*{07mm}%
\raisebox{0.25\columnwidth}{%
\begin{minipage}[t]{0.5\columnwidth}
\small\raggedright
Parameters:
{\bf (a)} Off-resonant, short damping memory: bare frequencies $\omega_{20} = 1.15\,\omega_{10}$, 
Ohm-Drude damping~(\ref{eq:Ohm-Drude-damping}) with $\gamma_1 = \gamma_2 = 0.1\,\omega_{10}$
and cutoff $1/\tau_c = 50\,\omega_{10}$. Coupling $\lambda / m_1 = 0.09\,\omega_{10}^2$
and $m_2 = m_1$. 
{\bf (b)}~Resonant, strong and different damping, critical coupling:
$\omega_{20} = \omega_{10}$, $\gamma_1 = 0.4\,\omega_{10}$, $\gamma_2 = 0.2\,\omega_{10}$,
$1/\tau_c = 50\,\omega_{10}$, coupling $\lambda / m_1 = 0.27\,\omega_{10}^2$.
{\bf (c)}~Resonant, long damping memory: $\omega_{20} = \omega_{10}$,
$\gamma_1 = 0.25\,\omega_{10}$, $\gamma_2 = 0.217\,\omega_{10}$,
narrow bandwidth $1/\tau_c = 2\,\omega_{10}$, 
coupling $\lambda / m_1 = 0.36\,\omega_{10}^2$.
--
Without loss of generality, $m_1 = m_2$ in all cases.
\end{minipage}%
}
\caption[]{Loss spectra for two coupled oscillators in different regimes.
Dashed lines: absorption spectrum for non-coupled oscillator (red = lower frequency
$\oMega_{10}$,
blue = higher frequency $\oMega_{20}$). Solid lines: absorption spectrum per drive power 
injected into the red (line 1) or blue (line 2) oscillator.
(a) Detuned oscillators, strong coupling, Markovian bath.
(b) Degenerate frequencies, different damping, close to critical coupling.
(c) Degenerate frequency, close damping, above critical coupling,
long-memory bath. The vertical lines give the lossless resonance frequencies
of the non-coupled (dotted) and coupled (solid) oscillators 
[see Eq.\,(\ref{eq:eigenfrequencies}) for the latter]. A Lorentzian
bath spectrum is assumed whose bandwidth is much larger than $\gamma_1$ for
(a) and (b).}
\label{fig:loss-spectra}
\end{figure}

This behaviour is illustrated in Fig.\,\ref{fig:loss-spectra}. We consider a typical
setting of ``absorption spectroscopy'' where two external
monochromatic forces with amplitudes $f_1$, $f_2$ and frequency $\omega$ perturb 
the two oscillators. This drives the oscillators out of equilibrium, but the external
energy is eventually dumped into the baths. In the long-time limit, 
the total absorbed power is given by (for more details, see 
Sec.\,\ref{s:heat-current-spectrum})
\begin{equation}
\overline{ \langle \dot x \rangle f_1 + \langle \dot y \rangle f_2 } 
= 
\oMega |f_1|^2 \mathop{\rm Im} \frac{ \Chi_2(\oMega) }{ D(\oMega) }
+
\oMega |f_2|^2 \mathop{\rm Im} \frac{ \Chi_1(\oMega) }{ D(\oMega) }
\label{eq:def-absorption}
\end{equation}
where the overline denotes the time average.
The imaginary part $\oMega \mathop{\rm Im} [ \Chi_2(\oMega) / D(\oMega) ]$
can also be interpreted as the spectral function (effective mode density) for the
oscillator~1. The dotted lines in the Figure correspond to the non-coupled oscillators 
where $\oMega \mathop{\rm Im} [ 1/\Chi_i(\oMega) ]$ peaks.
The vertical solid
lines give the normal mode frequencies $\omega_{\pm}$~[Eq.\,(\ref{eq:eigenfrequencies})].
For detuned oscillators, the absorption spectra show a peak and a shoulder 
at the eigenfrequencies [panel (a)]. At critical coupling, the normal mode
splitting is barely larger than the oscillators' linewidths [panel (b)].
A colored bath [panel (c), see Eq.\,(\ref{eq:Lorentzian-bath})]
shifts the absorption peaks significantly relative to the 
prediction~(\ref{eq:eigenfrequencies}).

In the preceding plots, we consider a memory kernel with a Drude regularisation. 
It corresponds to the spectral density
\begin{equation}
\rho( \oMega ) = 
\frac{ \gamma / \tau_c^2 }{ \oMega^2 + 1/\tau_c^2 }
\label{eq:Lorentzian-bath}
\end{equation}
where $\gamma$ gives the overall scale of the friction coefficient and $\tau_c$ sets
the bath correlation time. Its friction kernel is a simple exponential given in
Appendix~\ref{s:Ohm-Drude}.

\subsection{Stationary covariances}
\label{s:correlations}

From the Fourier relation~(\ref{eq:linear-response-matrix}), we can construct
the stationary behavior of the coupled oscillators. One remarks first of
all that the poles of the response matrix that define the eigenfrequencies
of the system, are located below the real axis so that the system loses
the memory of its initial conditions (damped oscillator). (The same would be true
for poles that coalesce into branch cuts for certain spectral densities.) The mean values
of the phase space coordinates $(x, p_x, y, p_y)$ thus vanish on time
scales larger than $\sim 1/\mathop{\rm Re} \mu_i(\oMega_i)$. The same
result is obtained by taking the mean value of Eq.\,(\ref{eq:linear-response-matrix}),
since the Langevin forces vanish on average.

In the following, we focus on the fluctuations around these mean values. They
are captured by the covariance matrix defined by analogy to
Eq.\,(\ref{eq:def-covariance}):
\begin{equation}
C_{ij} = \lim_{t\to\infty} \langle q_i(t), q_j(t) \rangle_{12}
\,,
\qquad
(q_i) = (x, p_x, y, p_y)^{\sf T}
\label{eq:def-covariance-matrix}
\end{equation}
where the double index $12$ reminds that the baths 
coupled to the two oscillators are taken in equilibrium
at temperatures $T_1$ and $T_2$.

We illustrate the calculation with the difference coordinate 
studied by Dorofeyev~\cite{Dorofeyev2013a} because it determines the
average interaction energy. The Fourier solution yields 
\begin{equation}
x(\oMega) - y(\oMega) 
= 
\frac{ \Chi_2(\oMega) - \lambda }{ D(\oMega) } F_1(\oMega) 
-
\frac{ \Chi_1(\oMega) - \lambda }{ D(\oMega) } F_2(\oMega) 
\label{eq:x-y-in-Fourier}
\end{equation}
When we compute the average $\langle \left(x(t) - y(t)\right)^2 \rangle_{12}$ from
the Fourier transform~(\ref{eq:x-y-in-Fourier}), the 
fact~(\ref{eq:result-corr-FF-Fourier}) that different frequencies are not
correlated, implies that the average is stationary. We thus drop the time
argument and find
\begin{eqnarray}
\langle (x - y)^2 \rangle_{12} &=& 
\int\limits_0^\infty\!\frac{ {\rm d}\oMega }{ 2\pi }
\left\{
\left|\frac{\Chi_2(\oMega) - \lambda}{D(\oMega)}\right|^2 S_{F1}( \oMega ) 
\right.
\nonumber\\
&& \qquad \left. {}
+
\left|\frac{\Chi_1(\oMega) - \lambda}{D(\oMega)}\right|^2 S_{F2}( \oMega ) 
\right\}
\label{eq:}
\end{eqnarray}
It can be checked that this formula agrees with Eqs.(12--14) of 
Ref.\,\cite{Dorofeyev2013a} where the covariance 
$- \lambda \langle x, y \rangle_{12}$ is given separately 
(mean interaction energy). By similar calculations, we get the 
diagonal element of the covariance matrix~(\ref{eq:def-covariance-matrix}) 
\begin{eqnarray}
C_{xx} &=& \int\limits_0^\infty\!\frac{ {\rm d}\oMega }{ 2\pi }
\frac{ |\Chi_2(\oMega)|^2 S_{F1}(\oMega) + \lambda^2 S_{F2}(\oMega) }{ |D(\oMega)|^2 }
\label{eq:Cxx-result}
\end{eqnarray}
and an analogous result for $C_{yy}$. 
We see here how the thermal spectrum
of bath~$2$ couples to the oscillator position fluctuations; this second term 
in Eq.\,(\ref{eq:Cxx-result}) is naturally proportional to the oscillator
coupling $\lambda$.
The mixed position-momentum 
$C_{xp_x}$ covariance vanishes. This result and the expression for
the mixed correlation $C_{xp_y}$ between the oscillators are discussed next.

\subsection{Crossed correlations}
\label{s:crossed-correlation-spectra}

One key insight of the quantum Langevin model \cite{Dorofeyev2013a} is 
that without much further
effort, the preceding results also provide the two-time correlation functions
of the oscillator fluctuations in the stationary state. This
state is not in thermal equilibrium: due to the difference in bath
temperatures, there is actually a heat current flowing through the
system. The cross-correlation functions thus provide insight into the 
dynamical aspects of deviations from a local equilibrium state. 

We start with a general expression for a two-time correlation between
observables $A$ and $B$
\begin{eqnarray}
&& \langle A(t), B(t') \rangle_{12}
\nonumber\\
&& =
\int\!\frac{ {\rm d}\oMega }{ 2\pi }\frac{ {\rm d}\oMega' }{ 2\pi }\,
{\rm e}^{ {\rm i} (\oMega t - \oMega' t') }
\sum_{ij} \alpha_i^*(\oMega) 
\beta_j(\oMega') 
\langle F_i^\dagger( \oMega ), F_j( \oMega' ) \rangle_{12}
\nonumber\\
&&=
\int\!\frac{ {\rm d}\oMega }{ 4\pi }\,
{\rm e}^{ {\rm i} \oMega (t - t') }
\sum_{i} \alpha_i^*(\oMega) 
\beta_i(\oMega) 
S_{Fi}( \oMega )
\label{eq:def-spectrum-AB}
\end{eqnarray}
Here, the functions $\alpha_i(\oMega)$ and $\beta_i(\oMega)$ provide
the response of the observables $A$ and $B$ to the Langevin forces $F_i$
of the baths, $A(\oMega) = \alpha_1(\oMega) F_1( \oMega ) + \alpha_2(\oMega) F_2( \oMega )$.
In the second line, the bath spectral densities
$S_{Fi}( \oMega )$ from Eq.\,(\ref{eq:result-corr-FF-Fourier}) are used.
From Eq.\,(\ref{eq:FD1-Langevin-force}), they are
symmetric in $\oMega$ so that positive and negative frequencies can be combined.
In this step,
the identities $\alpha_i^*(\oMega) = \alpha_i(-\oMega)$ are useful
which must hold for a hermitean observable $A(t)$. 
This leads to the following convention for the 
\emph{cross-correlation spectrum} $S_{AB}( \oMega )$ (in general a complex quantity)
\begin{eqnarray}
\langle A(t), B(t') \rangle_{12} &=& 
\int\limits_0^\infty\!\frac{ {\rm d}\oMega }{ 2\pi }\,
\mathop{\rm Re}
\big[
{\rm e}^{ {\rm i} \oMega (t - t') }
S_{AB}( \oMega )
\big]
\label{eq:def-spectrum-S-AB}
\\
S_{AB}( \oMega )
&=&
\sum_{i} 
\alpha_i^*(\oMega) 
\beta_i(\oMega) 
S_{Fi}( \oMega )
\label{eq:general-spectrum-AB}
\end{eqnarray}

As an example, consider the pair $(A,B) = (x, p_x)$. We can read off
$\alpha_i(\oMega)$ ($i = 1,2$) from the first line of the linear response
matrix~(\ref{eq:linear-response-matrix}). In addition, from $p_x = m_1 \dot x$
follows the simple
relation $\beta_i(\oMega) = - {\rm i} m_1 \oMega \alpha_i(\oMega)$.
The correlation spectrum becomes
\begin{eqnarray}
S_{xp_x}( \oMega )
&=&
- {\rm i} m_1 \oMega
\frac{ |\Chi_2|^2 S_{F1} + \lambda^2 S_{F2} }{ |D|^2 }
\end{eqnarray}
where the frequency arguments are suppressed for simplicity.
This is purely imaginary and therefore, the equal-time
correlation $\langle x, p_x \rangle$ vanishes.

The results for the cross-covariance matrix can be collected into a matrix
(boldface for $2\times2$-matrices)
\begin{equation}
{\bf C}(t - t') = 
\begin{pmatrix}
\langle x(t), y(t') \rangle_{12} & \langle x(t), p_y(t') \rangle_{12}
\\
\langle p_x(t), y(t') \rangle_{12} & \langle p_x(t), p_y(t') \rangle_{12}
\end{pmatrix}
\label{eq:def-correlations-C}
\end{equation}
and have the following cross-correlation spectra
\begin{equation}
\begin{pmatrix}
S_{xy} & S_{xp_y}
\\
S_{p_xy}, & S_{p_xp_y}
\end{pmatrix}
= 
\lambda
\begin{pmatrix}
1 & - {\rm i} m_2 \oMega \,
\\
{\rm i} m_1 \oMega & m_1 m_2 \oMega^2
\end{pmatrix}
\frac{ K_2^* S_{F1} + K_1 S_{F2} }{ |D|^2 }
\label{eq:cross-correlation-spectra}
\end{equation}
The last fraction becomes real in the case of equal temperatures because
Eq.\,(\ref{eq:re-mu-and-DOS}) yields
\begin{equation} 
{\rm i} \oMega \left[
\mu_2^* \rho_1 
-
\mu_1 \rho_2
\right]
= \mathop{\rm Im}(\mu_2) \rho_1 
+ \mathop{\rm Im}(\mu_1) \rho_2
\label{eq:}
\end{equation}
so that the mixed position-momentum correlations vanish at equal times,
$\langle x, p_y \rangle_{12} = 0 = \langle p_x, y \rangle$.

\subsection{Heat current spectrum}
\label{s:heat-current-spectrum}

\begin{figure}[htbp]
\centerline{%
\includegraphics*[width=0.4\columnwidth]{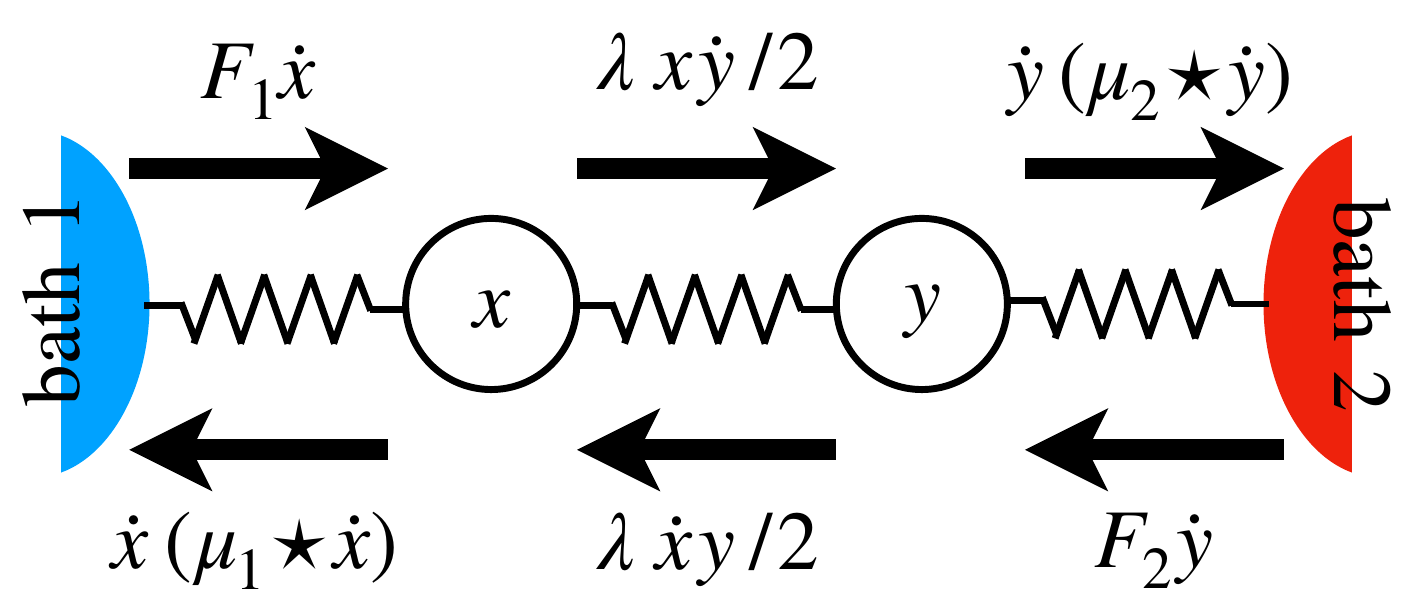}}
\caption[]{Illustration of power exchanged between oscillators and
baths. The net heat current is given by the difference
$\dot Q_{1\to2} 
= \frac{1}{2} \lambda \, \langle x \dot y - \dot x y \rangle_{12}$
where the average is taken in the non-equilibrium stationary state,
assuming two different temperatures for the heat baths.}
\label{fig:heat-balance}
\end{figure}

The correlation function 
$\langle x(t), p_y(t') \rangle$ actually captures the heat current 
through the link between the oscillators. To see this, we calculate
the power exchanged by oscillators and heat baths (see Fig.\,\ref{fig:heat-balance}).
It follows from the time-averaged derivative of the energy of the (isolated)
oscillator $(x, p_x)$
\begin{equation}
\overline{ \Big\langle \frac{ {\rm d} H_1 }{ {\rm d}t } \Big\rangle } 
= 
\lambda \, \overline{ \langle \dot x ( y - x ) \rangle }
- \overline{ \langle \dot x (\mu_1 \!* \dot x) \rangle }
+ \overline{ \langle F_1 \dot x \rangle }
\label{eq:energy-balance-1}
\end{equation}
The first term is the power
transferred by the connecting spring. 
The second term is negative definite (for any friction kernel)
and can be interpreted as the power dissipated into heat bath~$1$.
Finally, the last term is the rate of work performed by the Langevin force
$F_1$ on the oscillator $(x, p_x)$. 

These assignments of energy fluxes are illustrated in Fig.\,\ref{fig:heat-balance}.
We have used that $\overline{ \langle \dot x x \rangle } = 0$ and
$\overline{ \langle \dot x y \rangle } + \overline{ \langle x \dot y \rangle } = 0$ 
in a stationary state
to re-write
the power exchanged by the oscillators in a more ``anti-symmetric'' way. 
The energy balance illustrates why the absorption spectrum defined in
Eq.\,(\ref{eq:def-absorption}) is equivalent to the power dissipated
into the bath (in linear response to the perturbing fields). 
In the stationary state, it justifies that 
the heat current from bath~1 to bath~2 across the two oscillators can be
computed as%
\cite{L-z}
\begin{equation}
\dot Q_{1\to2} = 
\frac{\lambda}{2} \, \overline{ \langle x \dot y - \dot x y \rangle }
\label{eq:def-heat-current}
\end{equation}
(Any other point in the chain would give the same value, of course.)
In the stationary state, we identify this with the correlation
function $\langle x, \, p_y \rangle_{12}/m_2 - \langle p_x, \, y \rangle_{12}/m_1$
and get from Eq.\,(\ref{eq:cross-correlation-spectra})
the following spectral representation:
\begin{eqnarray}
\dot Q_{1\to2}
&=& \lambda^2 \!\int\limits_0^\infty\!\frac{ {\rm d}\oMega }{ 2\pi }
\oMega
\frac{ S_{F1} \mathop{\rm Im} \Chi_2^* 
- 
S_{F2} \mathop{\rm Im} \Chi_1^* }{ |D|^2 }
\label{eq:}
\end{eqnarray}
This simplifies with Eqs.(\ref{eq:FD1-Langevin-force}, \ref{eq:def-chi}) to
\begin{eqnarray}
\dot Q_{1\to2}
&=& 4 \lambda^2 \!\int\limits_0^\infty\!\frac{ {\rm d}\oMega }{ 2\pi }
\oMega^2 
\frac{ \rho_1 \rho_2 }{ |D|^2 }
\left(
 \vartheta_1 -  \vartheta_2 
\right)
\label{eq:result-heat-current-spectrum}
\end{eqnarray}
where the relation~(\ref{eq:re-mu-and-DOS}) between damping kernel and 
spectral density was used.
The net heat current vanishes when both baths are at the same temperature,
and heat flows from the hot to the cold bath because the spectral densities
$\rho_i(\omega)$ are non-negative.

\begin{figure}[htbp]
\centerline{%
\includegraphics*[width=0.45\columnwidth]{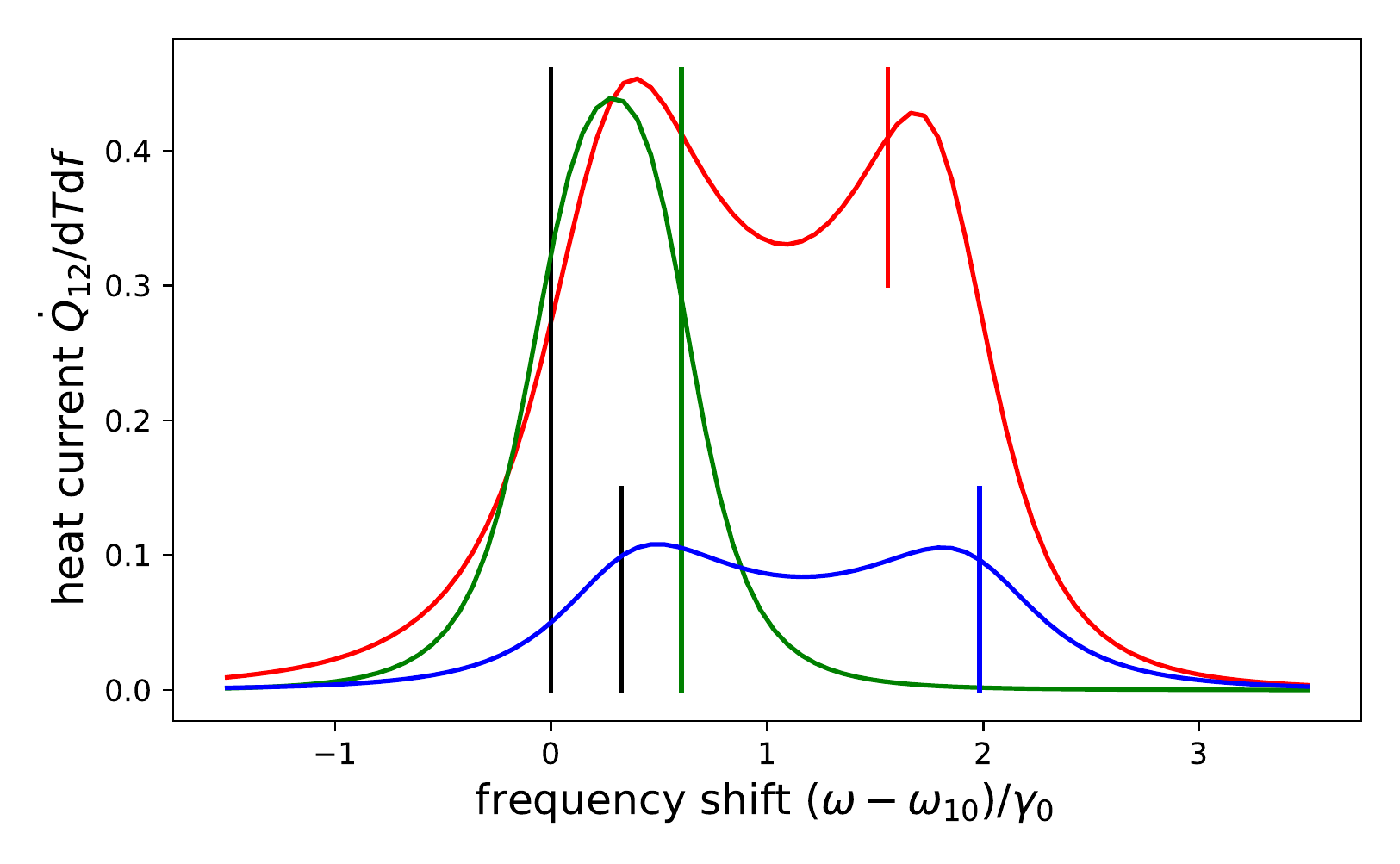}
\includegraphics*[width=0.45\columnwidth]{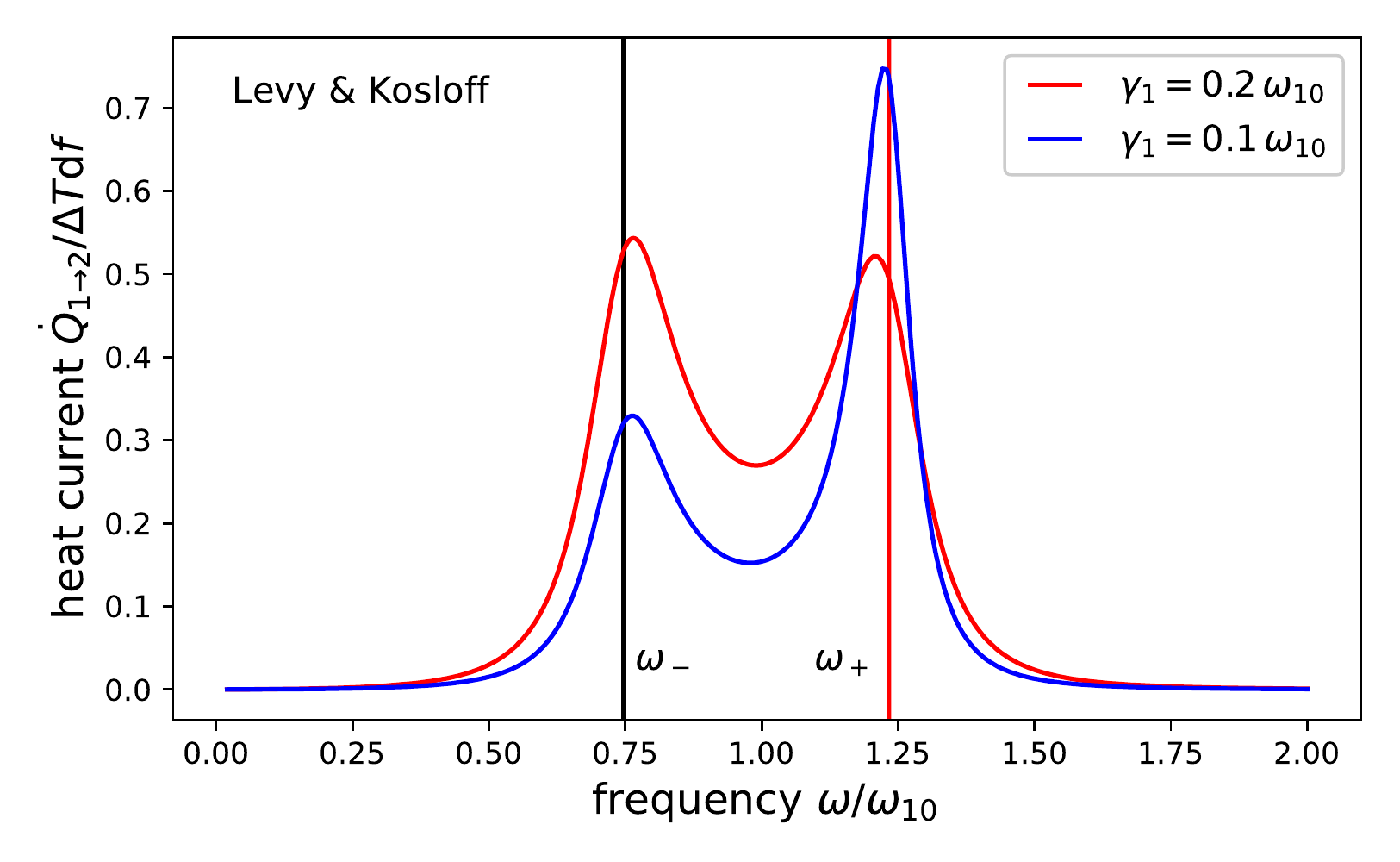}
}
\caption[]{(\emph{left}) Spectrum of heat current for small temperature difference
${\rm d}T \ll \tfrac12(T_{1} + T_{2}) = \hbar\oMega_1$. The curves labelled
(a), (b), (c) correspond to the cases shown in Fig.\,\ref{fig:loss-spectra}.
The vertical lines denote the normal mode frequencies from 
Eq.\,(\ref{eq:eigenfrequencies}). In case~(b), they are not resolved due
to strong damping.
(\emph{right}) Heat current spectrum for parameters close to Levy \& Kosloff
\cite{Levy2014}: 
different temperatures $T_1 = 5\,\hbar\omega_{10} >
T_2 = 4\,\hbar\omega_{10}$ with $\omega_{10} / T_1 > \omega_{20} / T_2$
(the spectrum is normalised by $\Delta T = T_1 - T_2$).
Detuned oscillators $\omega_{20} = 0.6\,\omega_{10}$,
identical dampings $\gamma_1 = \gamma_2$ as given in inset, memoryless bath
$1/\tau_c = 50\,\omega_{10}$, strong coupling $\lambda/m_1 = 0.36\,\omega_{10}^2$.
}
\label{fig:heat-spectrum}
\end{figure}

The heat current spectrum is illustrated in Fig.\,\ref{fig:heat-spectrum}.
On the \emph{left}, we consider a differential heat gradient $T_1 = T_2 + {\rm d}T$
where
$
 \vartheta_1(\oMega) -  \vartheta_2(\oMega) =
 {\rm d}T (\hbar\oMega/ T)^2 \bar{n} ( \bar{n} + 1 )
$
with the Bose-Einstein distribution $\bar{n} = \bar{n}( \oMega )$ evaluated 
at the mean bath temperature $T$. We plot the dimensionless quantity 
${\rm d}\dot Q_{12} / {\rm d}T {\rm d}f$ which is defined as the integrand 
of Eq.\,(\ref{eq:result-heat-current-spectrum}) (${\rm d}f = {\rm d}\omega/2\pi$).
For detuned oscillators, the eigenfrequencies
define peaks of efficient heat transport [Fig.\,\ref{fig:heat-spectrum}(\emph{left}), case (a)]. 
Note that in the critical coupling scenario (b),
only one peak is visible because the splitting of the normal modes is still
comparable to the linewidth. Strong coupling separates the two channels
in frequency [case (c)].
When comparing to the absorption spectrum of Fig.\,\ref{fig:loss-spectra},
it is interesting that the latter is larger in case~(a), although the heat current
stays relatively weak. This may be attributed to the small overlap
between the two resonances.

The motivation for Fig.\,\ref{fig:heat-spectrum}(\emph{right}) with a finite
temperature difference $\Delta T = T_1 - T_2$ is the comparison 
to Levy \& Kosloff \cite{Levy2014}.
The heat current given by Eq.\,(\ref{eq:result-heat-current-spectrum}) is one 
of the main results of the present paper
because Clausius' formulation of the second law of thermodynamics 
is satisfied for a relatively wide
class of harmonic models. Even before the integration, the heat current has 
a positive definite spectrum
because except for the difference $\vartheta_1 - \vartheta_2$, the 
integrand in~(\ref{eq:result-heat-current-spectrum})
has a definite sign. This holds for any spectral densities $\rho_{1,2}$, be they
``structured'' (non-Markovian case) or flat.
Our result differs strongly from the model of
Ref.\,\cite{Levy2014} that also considered two oscillators coupled to
separate (local) baths and 
found for certain choices of parameters
a violation of the second law. In that formulation, the
heat current is proportional to 
\begin{equation}
\mbox{Ref.\,\cite{Levy2014}}: \qquad
\dot Q_{1\to2} \propto
{\rm e}^{ \hbar \omega_{20} / T_2 } 
- {\rm e}^{ \hbar \omega_{10} / T_1 } 
\label{eq:LevyKosloff}
\end{equation}
which is negative when $\omega_{10} / T_1 > \omega_{20} / T_2$ [the parameters taken
in Fig.\,\ref{fig:heat-spectrum}(\emph{right})].
The fact that the
oscillator frequencies appear here outside any frequency integral may be traced
back to the assumption that the coupling is based on the rotating-wave 
approximation~(\ref{eq:rwa-coupling}) that preserves the total occupation number
of the two oscillators. This is, however, a poor approximation when the eigenfrequencies
differ significantly. The example presented here thus demonstrates that 
a non-equilibrium steady state consistent with thermodynamics can be constructed
even with locally coupled baths. 
The anomalous behaviour in Ref.\,\cite{Levy2014}
is not likely due to the difference between local and global
couplings (in the latter case, the baths couple to the normal modes of the coupled
oscillators at their shifted eigenfrequencies), 
nor to the (non)Markovian character of the master equation (the present 
model is consistent whatever the memory of the friction kernels). 

We may only speculate how an inverted heat current could appear in the present framework.
A problem may arise from UV divergences when the momenta are coupled 
[see Eq.\,(\ref{eq:rwa-coupling})]
since that involves 
an additional factor $\omega^2$ compared to the coordinate coupling.
One can also think of certain renormalisation schemes that operate a subtraction 
in the spectral densities $\rho_{1,2}(\oMega)$, but then one should rather re-consider the
physical meaning of the subtraction for the non-equilibrium problem. (The subtracted
modes are a way to take into account a renormalised oscillator mass, for example.
\cite{WeissBook})

\subsection{Fluctuation-dissipation relations}
\label{s:FD-relations}

Before discussing the correlations in the non-equilibrium stationary state with 
respect to entanglement between the oscillators, 
we point out that in the case of equal bath temperatures, the correlation
functions satisfy fluctuation-dissipation (FD) relations. This result is
satisfying and perhaps not obvious because a canonical equilibrium state
is imposed for the two heat baths alone, while the oscillators' state is
reached dynamically by solving the equations of motion.
There are indeed system-bath models in
the literature that have been criticised for yielding stationary states
that do not conform with the canonical equilibrium state 
\cite{GardinerBook, Ford1996, Stockburger2016}.

When the coupled oscillators are in a global equilibrium state with
effective temperature $\vartheta( \oMega )$, the fluctuation--dissipation relation 
for the (complex) cross-correlation
spectrum according to the convention~(\ref{eq:def-spectrum-S-AB}) reads
\begin{equation}
S_{ij}( \oMega ) = -2 {\rm i} \frac{  \vartheta( \oMega ) }{ \oMega }
\left[
R_{ji}( \oMega ) -
R_{ij}^*( \oMega ) 
\right]
\,, \quad i, j = x, y
\label{eq:FD2-canonical-variables}
\end{equation}
Here $\vartheta( \oMega )$ depends only on one temperature.
The indices $i$, $j$ enumerate the oscillator coordinates
$x$, $y$, and the set of linear response functions $R_{ij}$ is defined by
\begin{equation}
\langle x( \oMega ) \rangle =
\sum_i R_{xi}( \oMega ) f_i( \oMega )
\label{eq:}
\end{equation}
They describe the response of the oscillator coordinate $x$ to an external 
force acting on oscillator $i$ (recall the absorption spectrum of Fig.\,\ref{fig:loss-spectra})
and are given for our model by the matrix
elements in Eq.\,(\ref{eq:linear-response-matrix}). Indeed, for this
linear system, the same response function applies for an external force 
and for the Langevin forces themselves. The only difference is that
for an external (classical) force, the average $\langle x(\oMega) \rangle$ is
nonzero.

\threesubsection{Proof of FD relation}
The correlation spectrum is a special case of the general 
correlation~(\ref{eq:general-spectrum-AB}). For $A = x_i$ and $B = x_j$,
the response functions are $\alpha_{k} = R_{ik}$ and
$\beta_{l} = R_{jl}$, so that (common argument $\oMega$ suppressed again)
\begin{eqnarray}
S_{ij} &=& 
\sum_k R^*_{ik} R_{jk} S_{Fk} 
=
4 \vartheta \sum_k R^*_{ik} R_{jk} \mathop{\rm Re}(\mu_k) 
\label{eq:proof-FDR-0}
\end{eqnarray}
using Eqs.(\ref{eq:FD1-Langevin-force}) with a common temperature
in the last step.
The response matrix $R_{ij}$ of Eq.\,(\ref{eq:linear-response-matrix})
is the inverse of $K_{ij}$, the matrix in Eq.\,(\ref{eq:Fourier-matrix})
that translates the equations of motion: $R_{ik} K_{kl} = \delta_{il}$
(summation over double indices). Take the complex conjugate of this
equation and multiply from the right with $R_{jl}$:
\begin{equation}
R_{ik}^* K_{kl}^* R_{jl} = R_{ji}
\label{eq:}
\end{equation}
We subtract from this relation the expression one gets by
multiplying $R_{jl} K_{lk} = \delta_{jk}$
from the left with $R_{ik}^*$, and get
\begin{equation}
R_{ik}^* \left( K_{kl}^* - K_{lk} \right) R_{jl} = 
R_{ji} - R_{ij}^*
\label{eq:proof-FDR}
\end{equation}
The matrix $K_{kl}$ on the left-hand side is symmetric [Eq.\,(\ref{eq:Fourier-matrix})] 
and its only elements with an imaginary part  
are the diagonal ones,
\begin{equation}
K_{kl}^* - K_{lk} = 2{\rm i} \oMega \mathop{\rm Re}(\mu_{k}) \, \delta_{kl} 
\label{eq:}
\end{equation}
Inserting this into Eq.\,(\ref{eq:proof-FDR}), we recognise the summand 
on the rhs of Eq.\,(\ref{eq:proof-FDR-0}), and elementary algebra gives
the fluctuation-dissipation relation~(\ref{eq:FD2-canonical-variables})
for the cross-correlation spectra $S_{ij}$. The FD relations of the baths is thus carried
over to the oscillator pair, provided the system is globally in equilibrium.

Three remarks are in order. (1) The correlation functions involving
momentum variables are easily dealt with using the equation of motion
$\dot x_i = p_i / m_i$. This gives just a multiplicative factor in Fourier 
space. (2) The fluctuation--dissipation relation does not need a 
weak-coupling assumption and is valid for arbitrary $\lambda$
\cite{WeissBook,Polevoi1975}. In our case, 
both the fluctuation spectrum and the response functions have their
poles shifted by the coupling between the oscillators, compared to the bare
oscillators.
(3) The Kubo--Martin--Schwinger (KMS) relations for correlation functions
are also satisfied by the present model if the two bath temperatures coincide.
They involve correlations with a fixed order of operators and state that
their spectra satisfy $\mathcal{S}_{AB}( \omega ) 
= {\rm e}^{ - \hbar \omega / T } \mathcal{S}_{BA}(\omega)$. This is obviously
related to detailed balance.
The KMS relations are used in the traditional proofs of the fluctuation--dissipation
relations. Sometimes, however, they can be used to \emph{define} thermal equilibrium
in pathological cases where the canonical ensemble fails because its partition
function diverges.

\section{Entanglement}

The concept of entanglement tries to identify and quantify correlations between the
oscillators (or more generally between two parts of a system) that cannot be
explained classically. Within the seminal discussion of Einstein, Podolsky and Rosen
\cite{Einstein1935},
such correlations suggest an incomplete, non-local, or non-realistic interpretation 
of the joint quantum state of two particles, adopting the language of 
Bell \cite{Bell1964}. Entanglement measures have been developed over the last 20 years
to quantify the amount of non-classical correlations, for example, using the
magnitude of violating a Bell inequality. In the context of oscillators, 
entanglement theory speaks of continuous variables (rather than qubits or
other finite-dimensional systems). See Ref.\,\cite{Eisert2003b} for a review
and original work by Duan et al.~\cite{Duan2000} and Simon \cite{Simon2000}.
Extensions to more than two oscillators were proven by Werner \cite{Werner2001}
and to non-Gaussian states in Refs.\,\cite{Hillery2006, Agarwal2005, Shchukin2005}.

\subsection{Covariances and optimal EPR correlations}
\label{s:EPR-correlations}

The entanglement between the two oscillators in the stationary state
may be characterised via their covariance matrix $\Cov = (C_{ij})$ defined
in Eq.\,(\ref{eq:def-covariance-matrix}), provided the state is Gaussian 
which is the case here \cite{Talkner1981}.
This is a real symmetric 
$4\times4$-matrix that we write in the chosen basis $(q_i) = (x, p_x, y, p_y)$ 
in block form
\begin{equation}
\Cov = \begin{pmatrix}
{\bf A} & {\bf C} \\
{\bf C}^{\sf T} & {\bf B}
\end{pmatrix}
\label{eq:def-C-in-block-form}
\end{equation}
The blocks ${\bf A}$ and ${\bf B}$
describe the covariances of oscillator $(x, p_x)$, resp. $(y, p_y)$,
while the matrix ${\bf C}$ describes correlations among the two oscillators.
Its two-time version was given
in Eq.\,(\ref{eq:def-correlations-C}).
The Duan--Simon criterion
\cite{Duan2000, Simon2000}
states that the two oscillators are in a separable (i.e.,
non-entangled) state if and only if the following inequality
is satisfied:
\begin{eqnarray}
\det( {\bf A} ) \det( {\bf B} ) +
\left( |\det( {\bf C} )| - \tfrac{1}{4}\hbar^2 \right)^2
- I_4
\nonumber\\
\ge
\tfrac{1}{4}\hbar^2 \left[ \det( {\bf A} ) + \det( {\bf B} ) \right]
\label{eq:Duan00-Simon00}
\end{eqnarray}
(We have adapted the formulation of Ref.\,\cite{Simon2000} to
dimensional positions and momenta.)
Here, $I_4$ is the fourth invariant of the covariance matrix $\Cov$ under
local canonical transformations (the other three are the determinants
in Eq.\,(\ref{eq:Duan00-Simon00}):
\begin{equation}
I_4 = \mathop{\rm tr}(\sigma {\bf A} \sigma {\bf C} \sigma {\bf B}
\sigma {\bf C}^{\sf T})
\label{eq:def-I4}
\end{equation}
where $\sigma$ is the so-called symplectic matrix
that collects the commutation relations among the phase-space coordinates
\begin{equation}
[q_i, q_j] = {\rm i} \hbar \sigma_{ij}
\label{eq:def-symplectic-matrix}
\end{equation}
In Eq.\,(\ref{eq:def-I4}), a $2\times2$ version of $\sigma$ is used. 
A linear coordinate transformation $q_i \mapsto Q_i = \sum_j \S_{ij} q_j$ on the
total phase space
is canonical if it preserves the commutation relations. Collecting $2\times2$ blocks
into the symplectic matrix $\sigma$, this is equivalent to
$\S \sigma \S^{\sf T} = \sigma$.
For details on linear canonical transformations
and the symplectic groups Sp(2), Sp(4) they form, see 
Refs.\cite{Arvind1995b, Eisert2003b, Adesso2004}. 

The physical meaning of the criterion~(\ref{eq:Duan00-Simon00}) is that
separable states never show Einstein-Podolsky-Rosen (EPR) 
correlations \cite{Duan2000}.
Recall that these correlations imply that for an entangled
(gaussian) state, there is a pair $(Q,P)$
of sum or difference variables
whose uncertainty product is below the Heisenberg limit, 
$\Delta Q \Delta P < \hbar/2$. This means
that a measurement
on one oscillator permits to infer the coordinates of the other one
(`EPR paradox'). 
In a symmetric situation, the EPR pair may be given by the combinations 
$Q = (x + y)/\sqrt{2}$ and $P = (p_x - p_y)/\sqrt{2}$.
The point is not that such a situation would really violate the Heisenberg 
relation ($Q$ and $P$ actually commute in this example),
but that for a separable state, the uncertainty product $\Delta Q \Delta P$ 
is bounded from below by $\hbar / 2$, as shown in Ref.\,\cite{Duan2000}.

For the entanglement criterion of Refs.\,\cite{Duan2000, Simon2000},
one diagonalises
the so-called partially transposed covariance matrix 
\begin{equation}
{\Cov}^{\Gamma} = \Gamma\, \Cov \,\Gamma
\label{eq:def-PT-covariance-matrix}
\end{equation}
Here, the partial transposition acts as
$\Gamma = \mathop{\rm diag}( 1, 1, 1, -1)$ in the basis
$( x, p_x, y, p_y )$ [time reversal on oscillator $(y, p_y)$ alone].
The diagonalisation amounts to constructing a canonical coordinate transformation $\S$ 
such that
$\S {\Cov}^{\Gamma} \S^{\sf T}$ is diagonal. This procedure is called
symplectic diagonalisation (see Appendix\,\ref{a:symplectic-diagonalisation}).
The smallest (symplectic) eigenvalue 
$\eta_{\rm min}$ of ${\Cov}^{\Gamma}$
provides the entanglement measure called logarithmic negativity 
\cite{Vidal2002, Adesso2004}
\begin{equation}
E_{12} = \log\frac{ \hbar }{ 2 \eta_{\rm min} }
\qquad
\mbox{if } \eta_{\rm min} < \hbar/2
\,,
\label{eq:def-entanglement-nu-min}
\end{equation}
In the case $\eta_{\rm min} \ge \hbar/2$, 
the two oscillators are separable ($E_{12} = 0$), 
although they may still be classically correlated.
That can be quantified by the quantum mutual information or correlation 
entropy \cite{Adesso2004, Lang2011}.
This entropy considers the
difference between the naive additive expectation for a composite system 
\begin{equation}
S_{12} = S_1 + S_2 - S_{\rm tot}
\label{eq:def-correlation-entropy}
\end{equation}
Here, entropies are computed according to von Neumann as 
$S_{\rm tot} = - \mathop{\rm tr}( \rho_{12} 
\log \rho_{12} )$, while $S_1$ and $S_2$ are based on the reduced density
matrices (tracing out the other oscillator). 
They coincide with the thermodynamic entropy 
in the thermal equilibrium state (canonical ensemble),
up to a scale factor $k_B$, but they are 
well-defined even out of thermal equilibrium. 
In the present example, $\rho_{12}$ is the state of the two oscillators with the heat
baths traced out.
The sign $S_{12} \ge 0$ of the mutual information
becomes plausible when we recall that $S_1$ and
$S_2$ are computed from reduced states that are missing correlations
between the systems. This is even true at the classical level.
If the composite system is in an entangled pure state, then
$S_{\rm tot} = 0$ and the reduced states are mixed so that $S_i > 0$.

The symplectic eigenvalues and vectors of the covariance matrix
come in pairs (Williamson theorem~\cite{deGosson2006}, Appendix\,\ref{a:symplectic-diagonalisation}). 
We would like to point out
that the eigenvectors corresponding to $\eta_{\rm min}$ 
provide a simple way to construct an optimal EPR pair $(Q, P)$ 
of canonical coordinates
\begin{equation}
Q = \sum_{i} ( \S \,\Gamma )_{1i} q_i 
\,, \quad
P = \sum_{i} ( \S \,\Gamma )_{2i} q_i 
\label{eq:def-EPR-pair}
\end{equation}
The matrix $\S$ is ordered such that the first two lines 
correspond to the eigenvalue $\eta_{\rm min}$ of $\Cov^\Gamma$. 
The first two diagonal 
elements of $\S {\Cov}^{\Gamma} \S^{\sf T}$ yield indeed
\begin{equation}
\langle Q, Q \rangle_{12} = \eta_{\rm min} = \langle P, P \rangle_{12}
\label{eq:EPR-variances}
\end{equation}
while $\langle Q, P \rangle_{12} = 0$. 
Here, we have used the fact that the transformation $(Q, P) \mapsto
(Q/r, rP)$ with $r \ne 0$ is canonical so that $Q$ and $P$ may be
normalised to have the same physical dimension [homogeneous to
$\sqrt{\hbar}$ in this paper]
and the same variance.
Note that the averages in Eq.\,(\ref{eq:EPR-variances}) are computed
with respect to the non-equilibrium steady state. The only effect
of the partial transposition $\Gamma$
in Eq.\,(\ref{eq:def-EPR-pair}) is that $Q$ and $P$
are no longer canonically conjugate:
$[ Q, P ] \ne {\rm i}\hbar$. This was already the case for the
symmetric EPR pair introduced above. We conclude that 
since $\langle Q, Q \rangle_{12} = \Delta Q^2$,
the two oscillators are EPR-correlated when 
$\Delta Q \Delta P = \eta_{\rm min} < \hbar/2$.

\begin{figure}[tbh]
\centerline{%
\includegraphics*[width=0.5\columnwidth]{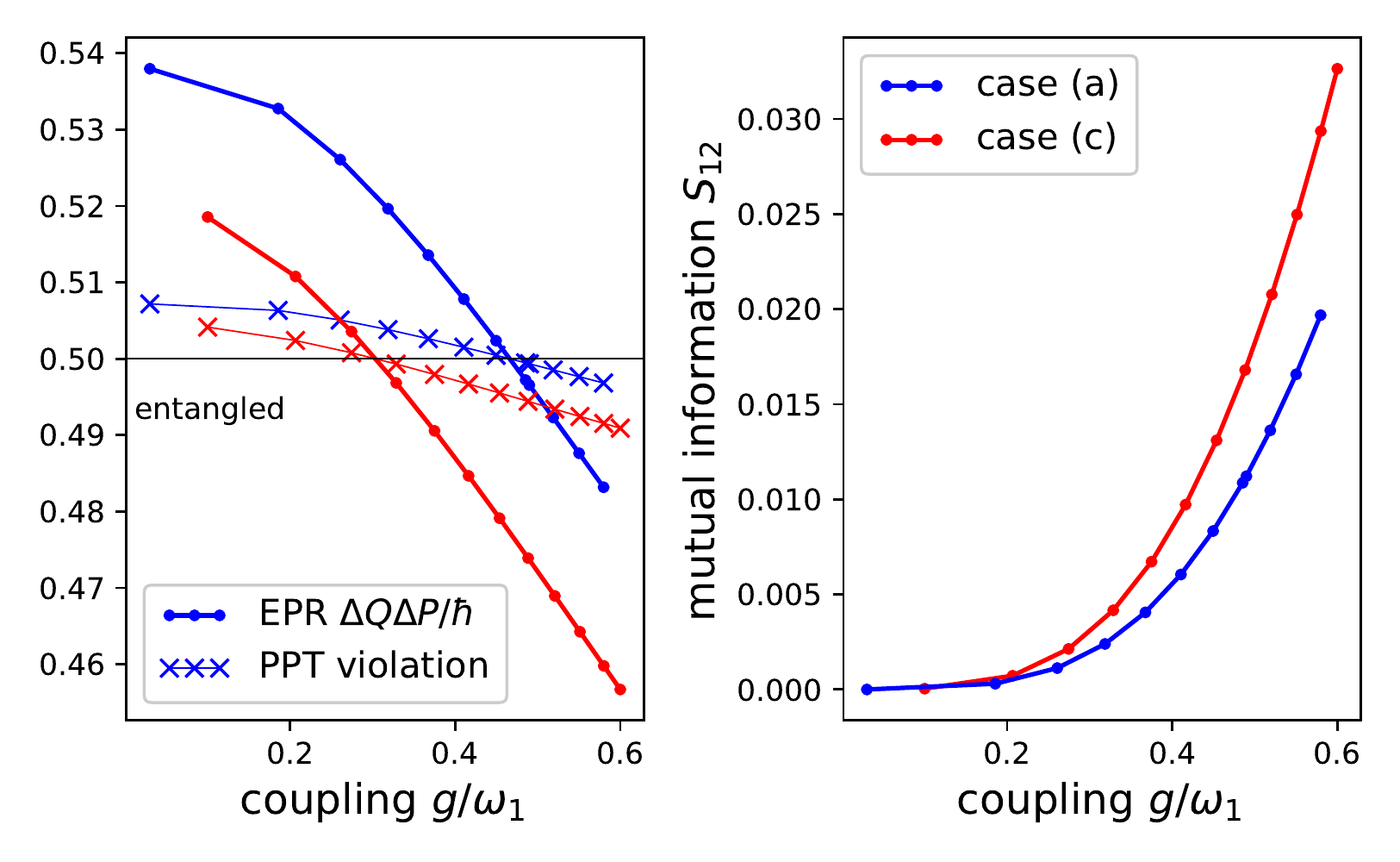}
}
\caption[]{(\emph{left})
Covariances that test the entanglement between the two oscillators
in the stationary state. Baths at low temperatures $T_1 = 0.1\,\hbar \oMega_{10}$,
$T_2 = 0.15\,\hbar \oMega_{10}$.
The curves marked (a, c) correspond to the
parameters of Fig.\,\ref{fig:loss-spectra},
only the coupling
between the oscillators is varied, expressed as
$g = (\lambda/m_1)^{1/2}$.
The crosses visualise the PPT criterion
for the partially transposed covariance matrix
$\Cov^\Gamma$: when the smallest (ordinary) eigenvalue of the hermitean 
matrix $\Cov^\Gamma + {\rm i} \hbar \sigma / 2$ falls below zero, the state
is entangled~\cite{Simon2000}. (Our plot shifts this eigenvalue up by $\hbar/2$
so that entanglement appears below the same line.)
The dots and curves give the uncertainty product $\Delta Q \Delta P$
of the EPR coordinate combinations~(\ref{eq:EPR-variances}) in units of $\hbar$, 
constructed from the smallest \emph{symplectic} eigenvalue of $\Cov^\Gamma$. 
(\emph{right})
Correlation entropy (mutual information) $S_{12}$ [Eq.\,(\ref{eq:def-correlation-entropy})]
vs. the coupling.
}
\label{fig:entanglement}
\end{figure}

A comparison of the EPR correlations found in this way and the logarithmic
negativity is shown in Fig.\,\ref{fig:entanglement} where the ``uncertainty
product'' $\Delta Q \Delta P$ of the EPR pair is shown (curves with dots)
for two parameter settings of Fig.\,\ref{fig:loss-spectra}, while
increasing the coupling $\lambda$. 
The data also show the mutual information~(\ref{eq:def-correlation-entropy}).
The crosses illustrate the so-called PPT
criterion: the oscillators are PPT-entangled when the matrix
$\Cov^\Gamma$ does not correspond to a physical state. 
(See Figure caption for more details.) The two criteria quantitatively
agree on the onset of entanglement (the curves cross there). 
Our construction of the EPR pair has
the advantage that it provides the experimenter with a definite measurement
protocol: choose the generalised coordinates $Q$, $P$ coming out of the
symplectic eigenvectors and measure their correlations.

The results of Fig.\,\ref{fig:entanglement} are 
for temperatures $T_1, T_2 \ll \hbar\oMega_1$.
We have evaluated the frequency integrals for the stationary covariance
matrix elements numerically. Baths with a Drude cutoff for the spectral
density are taken so that even the momentum correlations are UV-convergent.
The diagonalisation of the covariance matrices $\Cov$ and $\Cov^\Gamma$ is
done using the symplectic techniques of Appendix~\ref{a:symplectic-diagonalisation}.
We present data for a non-equilibrium situation, but no qualitative
changes appear when the two baths have similar temperatures. 

One sees that low temperatures and large couplings lead to both correlations and 
entanglement. Off-resonant oscillators requires a larger coupling (to provide an
efficient mixing in the normal modes).
Entanglement appears when the chosen covariances
are below the $\hbar / 2$ threshold (horizontal line).
Note that the mutual information $S_{12}$ is not in a one-to-one correspondence
with entanglement -- this is related to the fact that the 
variances $\Delta Q \Delta P$ (the eigenvalues $\nu$ of the covariance
matrix) are compared to the `quantum scale' $\hbar$
when dealing with entanglement. 
It is also remarkable that the EPR pair
constructed above yields an `uncertainty product' that falls, for entangled
oscillators, even below the PPT criterion (marked by crosses). 
The two entanglement quantifiers coincide, in fact, for the special case that the
two oscillators are in a so-called symmetric state (i.e., ${\bf A}$ and ${\bf B}$ 
have the same sympletic eigenvalues \cite{Adesso2004}), which is generally not
the case here. In the plot, it is seen that the quantifiers coincide 
right at the entanglement threshold. This is proven analytically
in Appendix~\ref{a:symplectic-diagonalisation}  
[after Eq.\,(\ref{eq:PPT-eigen-problem})].

The procedure of symplectic diagonalisation, although unfamiliar, also helps
in evaluating the mutual information $S_{12}$ of Eq.\,(\ref{eq:def-correlation-entropy}).
The symplectic eigenvalues $\nu_{\pm}$ of the covariance matrix
$\Cov$ itself quantify, loosely speaking, the occupation of the normal modes
$\omega_{\pm}$. Indeed, they 
determine the von Neumann
entropy of the Gaussian two-oscillator state as the simple sum \cite{Holevo2001}
(Eq.\,(23) of Ref.\,\cite{Adesso2004})
\begin{align}
S_{\rm tot} &= f( 2 \nu_+ / \hbar ) + f( 2 \nu_- / \hbar )
\label{eq:def-entropy-S12}
\\
f( x ) &= 
  \frac{ x + 1 }{ 2 } \log\frac{ x + 1 }{ 2 }
- \frac{ x - 1 }{ 2 } \log\frac{ x - 1 }{ 2 }
\label{eq:entropy-and-sympl-eigenvalue}
\end{align}
Note that Eq.\,(\ref{eq:entropy-and-sympl-eigenvalue}) only 
makes sense for $\nu_{+,-} \ge \hbar/2$: this is the criterion for 
a physical state.
A partial entropy like $S_1$ 
in Eq.\,(\ref{eq:def-correlation-entropy})
is easily computed from the covariance matrix because the reduced states
are determined by the block matrices ${\bf A}$ and ${\bf B}$. (Tracing out
the other oscillator amounts to chopping off unobserved blocks 
of the covariance matrix.)
For a
single oscillator, the symplectic diagonalisation transforms ${\bf A}$ 
into a multiple of the unit matrix.
We then get its symplectic eigenvalue $\nu_A$
from the determinant $\det{\bf A} 
= \det\left( {\bf S} {\bf A} {\bf S}^{\sf T} \right)
= \nu_A^2$,
and $S_1 = f( 2 \nu_A / \hbar )$.
(Linear canonical transformations have $\det {\bf S} = 1$.)

\subsection{Minimum noise quadratures in the frequency domain}

We now combine the two methods of our analysis employed so far
and introduce
ways to quantify entanglement via frequency spectra that are 
available for the non-equilibrium state of the two oscillators
(Sec.\,\ref{s:crossed-correlation-spectra}). The criteria for entanglement
discussed so far provide an obvious motivation: the two oscillators
are entangled when certain joint measurements (involving linear
combinations of observables) show errors below the ground state
uncertainties imposed by the Heisenberg relations. 
This suggests to analyse the spectrum $S_{QQ} = \frac{1}{2} \left(
S_{xx} + S_{yy} + 2 S_{xy} \right)$
of the EPR variance $Q = (x + y)/\sqrt{2}$ introduced above. But why prefer 
$S_{PP}$ to $S_{YY}$ with $Y = (x - y)/\sqrt{2}$? 
The two are related since, as noted above, 
the time derivative introduces just an additional factor $\oMega^2$ into the spectrum. 
And what would be the ``entanglement threshold'' for such a spectrum?

The spectral representation of the correlation matrix $\Cov = (C_{ij})$ 
gives us a matrix $S_{ij}(\oMega)$ that one may analyse for its local symplectic
invariants. This is not of much use, however:
from the block form for the spectrum of the off-diagonal sub-matrix ${\bf C}$ 
[see Eqs.(\ref{eq:cross-correlation-spectra}, \ref{eq:def-C-in-block-form})],
it is easy to check that at any frequency $\omega$, this sub-determinant vanishes.
The same is true for the spectra that give the sub-blocks ${\bf A}$ and ${\bf B}$.
Another quick calculation checks that the formula~(\ref{eq:def-I4})
for the fourth invariant, when applied to the sub-blocks of the
spectral matrix $S_{ij}(\oMega)$, yields $I_4 = 0$, too. Whatever quantity
should replace the term $\hbar^2$ in Eq.\,(\ref{eq:def-I4})
in the spectral domain: the inequality~(\ref{eq:Duan00-Simon00}) is satisfied 
as long as it is real.

To simplify the following discussion, it is convenient to scale the 
canonical coordinates $(x, p_x, y, p_y)$ in such a way that they have
the same physical dimension. 
Recall that such a re-scaling
was already used when we introduced the EPR pair $Q$ and $P$ with identical
variances. 
The transformation in the $(x, p_x)$
plane we adopt is 
\begin{equation}
\begin{pmatrix}
x 
\\
p 
\end{pmatrix}
\mapsto
\begin{pmatrix}
\sqrt{k_1 / \oMega_{10}} & 0
\\
0 & 1 / \sqrt{m_1 \oMega_{10}}
\end{pmatrix}
\begin{pmatrix}
x 
\\
p 
\end{pmatrix}
\label{eq:}
\end{equation}
which is canonical if we set $\oMega_{10} = \sqrt{ k_1 / m_1 }$.
Note that we work here with the
convention of re-scaling with the \emph{non-coupled} eigenfrequencies.
This is motivated by experimental approaches where two parties would
have access to only one of the two oscillators, e.g., the parameters
$k_1, m_1$ and the canonical pair $(x, p_x)$. Simple consequences are 
that for the time derivative, we now have $\dot x = \omega_{10} p_x$,
and the (bare) oscillator energy is $H_1 = \omega_{10} ( p^2 + x^2 )$.

To search for spectral correlations, consider now a pair of coefficients 
$\alpha_x$, $\alpha_p$ and form the linear
combination
$a = \alpha_x \,x + \alpha_p \,p$. Switching to Fourier space, we have
\begin{equation}
a(\oMega) = \alpha_x \, x(\oMega) - {\rm i} (\oMega/\omega_{10}) \alpha_p \, x(\oMega)
\label{eq:}
\end{equation}
The real coefficients
thus combine in a natural way into a complex weight factor
$\alpha = \alpha_x + {\rm i} (\omega/\omega_{10}) \alpha_p$, as it happened
when the general cross-correlation spectrum $S_{AB}$ was introduced 
[Eq.\,(\ref{eq:general-spectrum-AB})]. When computing the spectrum 
$S_{aa}$, the global phase of $\alpha$ drops out (because we focus 
on a stationary state). This motivates to consider
the following hermitean spectral matrix [taking $A, B = x, y$
in Eq.\,(\ref{eq:general-spectrum-AB})] 
\begin{equation}
{\bf S}( \oMega ) = 
\begin{pmatrix}
S_{xx}( \oMega ) & S_{xy}( \oMega )
\\
S_{yx}( \oMega ) & S_{yy}( \oMega )
\end{pmatrix}
\label{eq:def-spectral-matrix}
\end{equation}
and to search for its (complex) eigenvectors in view of minimising covariance
spectra. Indeed, 
a complex linear combination $q(\oMega) = \alpha \, x(\oMega) + \beta \, y(\oMega)$ 
of the two oscillator
coordinates
would show a spectrum 
$S_{qq}( \oMega )
= (\alpha^*, \beta^*) {\bf S}( \oMega ) (\alpha, \beta)^{\sf T}$. 
This quadratic form can be minimised by choosing $(\alpha, \beta)$ as eigenvectors
of ${\bf S}$, that form the unitary matrix $U$ in the
similarity transformation 
${\bf S} \mapsto U^\dag {\bf S} U = \mathop{\rm diag}( S_{\rm min}, S_{\rm max})$.

We remark that such a construction can be understood in the time domain as a
``linear filter'' that maps the time series data $\{ x(t), y(t) \}$ to
\begin{equation}
q(t) = \int\!{\rm d}\tau\,
\Big\{
\alpha(\tau) \, x( t - \tau)
+ \beta(\tau) \, y( t - \tau)
\Big\}
\label{eq:def-linear-filter}
\end{equation}
This convolution takes in Fourier space the simple form 
$q(\Omega) = \alpha(\Omega) \, x(\Omega) + \beta(\Omega) \, y(\Omega)$ and suggests certain
constraints on the frequency-dependent coefficients $\alpha(\Omega)$ and $\beta(\Omega)$
(related to reality and causality). Such constraints may be relaxed, however, if one
considers that the linear filter is applied to pre-recorded data measured by the
experimenters who share their locally obtained data.

\begin{figure}[bth]
\includegraphics[width=0.45\columnwidth]{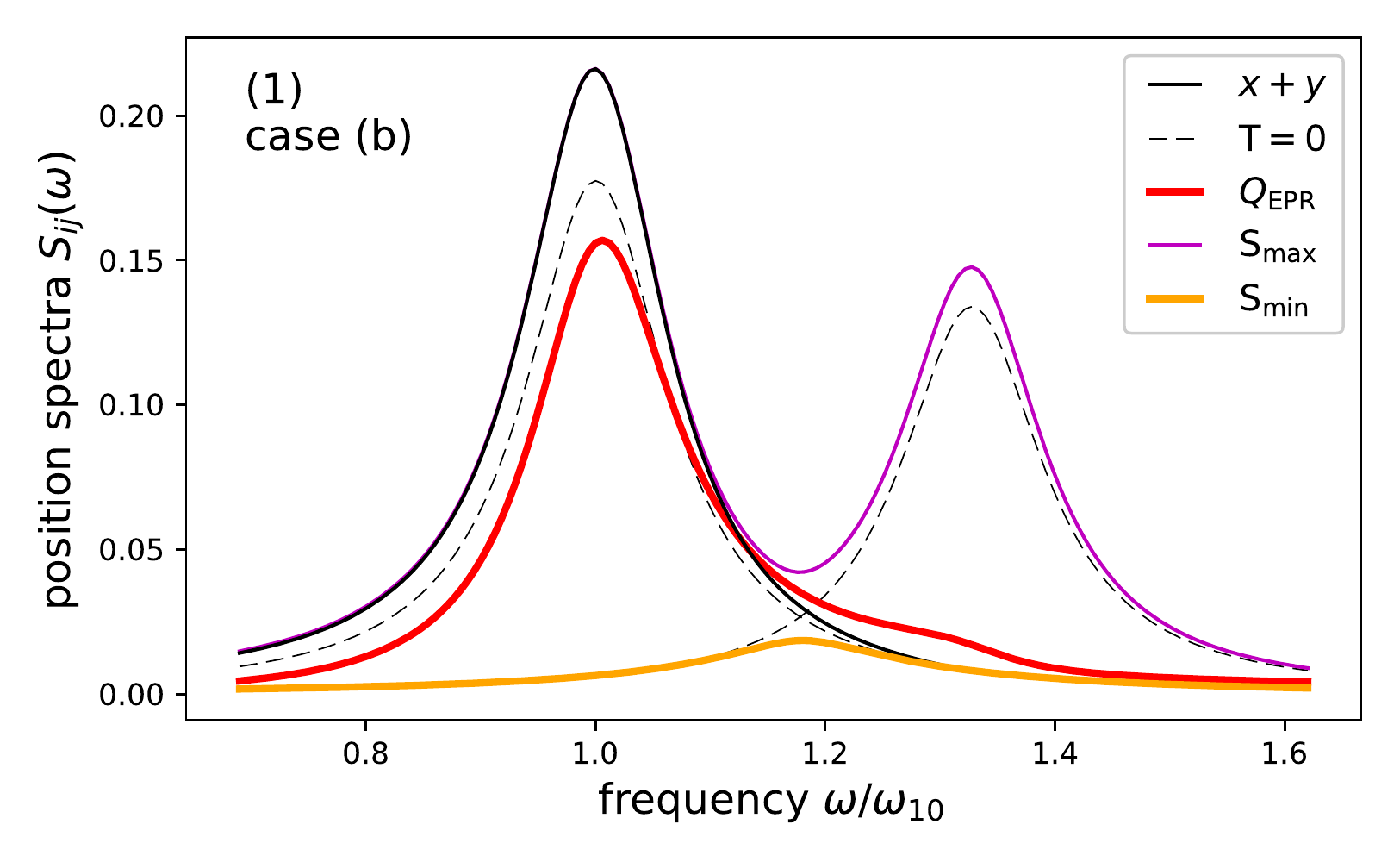}
\hspace*{02mm}
\includegraphics[width=0.45\columnwidth]{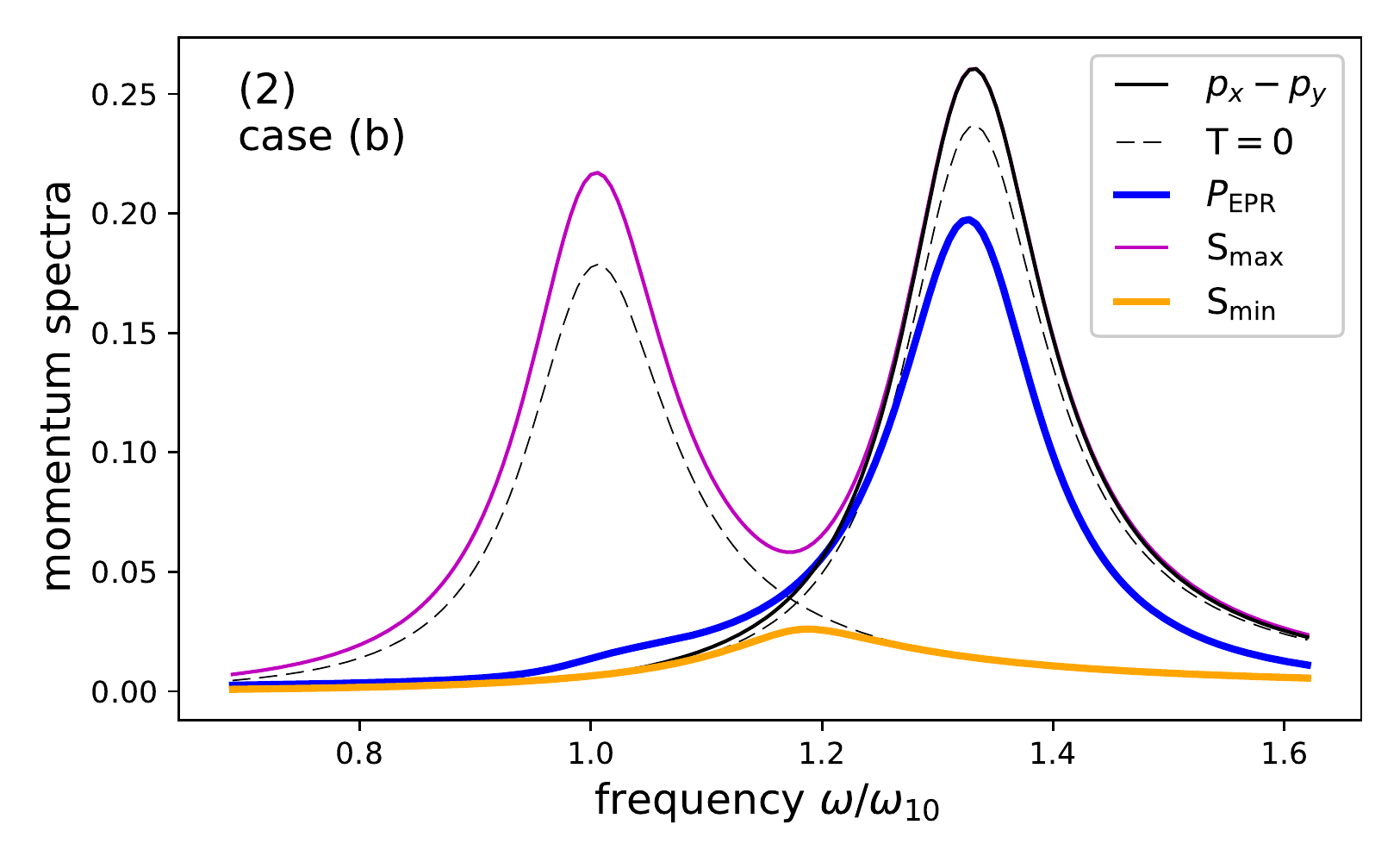}%
\caption[]{Spectra of optimal covariance witnesses. (\emph{left}) position
variables, (\emph{right}) momentum variables. Thin dashed lines: $T = 0$ limit
for the variables $(x \pm y)/\sqrt{2}$ and $(p_x \pm p_y)/\sqrt{2}$.
``EPR'': spectra for the fixed canonical coordinates $Q$, $P$ constructed
in Sec.\,\ref{s:EPR-correlations} 
that
minimise the partially transposed covariance matrix. Magenta and orange:
largest ($S_{\rm max}$) and smallest ($S_{\rm min}$) eigenvalue of the (hermitean) spectral matrix ${\bf S}(\oMega)$ [Eq.\,(\ref{eq:def-spectral-matrix})].
Parameter set (b) of Fig.\,\ref{fig:loss-spectra} in both panels,
$T_1 = 0.5\, \hbar\omega_{10} > T_2 = 0.25\, \hbar\omega_{10}$.
}
\label{fig:entanglement-spectra}
\end{figure}

The result of this construction, per each frequency,
is shown in the curves marked $S_{\max}$ and $S_{\min}$ in
Fig.\,\ref{fig:entanglement-spectra} where the largest (smallest)
of the two eigenvalues is plotted. It is to be noted that the spectrum
$S_{\min}$ is close to or even below the lower of two ``quantum limits'' that 
we construct in the following way: consider the sum and difference quadratures
$(x \pm y)/\sqrt{2}$ and compute their noise spectrum in the presence
of all couplings, but at temperature $T = 0$ for both baths. These
spectra show separate peaks at the normal mode frequencies. This is as 
expected, since in the lower (higher) normal mode the two oscillators move
in phase (in phase opposition). It is remarkable that the EPR coordinates
constructed from the ``global'' covariance matrix $\Cov$ already achieve
for certain frequencies a noise spectrum below this limit 
[red and blue curves in Fig.\,\ref{fig:entanglement-spectra}].

To conclude this discussion, we remark that the model presented here, being
based on the assumption that the two baths are not correlated, may be considered
as a kind of ``separable reference'' with respect to more general (non)classical
correlations. It is well known that two systems that couple to the \emph{same} bath
experience stronger correlations since the polarisation of the bath translates
into additional interactions (for example, the van der Waals interaction in atomic
and molecular physics)
\cite{Paz2009, Xiang2009, BuhmannBook_vol1}.
The minimal correlation spectra can thus 
provide a reference: if arbitrary correlations between
the Langevin forces are allowed for, one could imagine to get smaller covariance
spectra (tighter EPR correlations). The presence of such correlations must be
inferred when experimental data fall
below the spectral minima constructed here.
This may be expected in a situation of strong coupling when
entanglement between the oscillators gets mapped via bath polarisation into
the joint quantum state of the baths.

\section{Conclusion}

Coupled oscillators provide a paradigmatic example of system and bath models that
can be analytically solved with the help of quantum Langevin equations, without any
Markov approximations. In this paper, we have shown that a composite system
whose parts are locally coupled to heat baths passes all thermodynamic consistency
tests in the stationary state, be it non-equilibrium (different bath temperatures)
or not. We have shown that the fluctuation--dissipation relations are satisfied
for any choice of bath spectral density, provided the two bath temperatures coincide.
If they differ, the heat current is found as a positive definite frequency integral
in accordance with the Second Law. The anomalous heat current fromm cold to hot bath
found in earlier work \cite{Levy2014}
is probably due to approximations applied in the modelling (rotating-wave approximation,
Lindblad dissipation operators appropriate for resonant interactions). The criticism
of the concept of locally coupled baths that arose in the wake of Ref.\,\cite{Levy2014}
does not seem justified in view of the results presented here.

The coupled oscillator model has the advantage that it can be applied flexibly to a
plethora of physical systems. The mechanical oscillators in force microscopy
and resonant electric circuits provide two typical examples where
both weak and strong damping is relevant. The phonon bath in a solid can be easily modelled
with a Debye spectral density that shows qualitatively different memory effects
compared to the Ohm-Drude case used here in the numerical examples. 
One may even think of very different
resonance frequencies as they appear in optomechanical systems (light from a laser 
cavity coupled to a mirror mounted on an oscillating membrane)
\cite{Vitali2003a,Kleckner2006,Vahala2006a,Miao2010,Rossi2018}, where the
standard interaction (via radiation pressure) is mapped to a bilinear form under
suitable approximations. Another hybrid system would be provided in atom chips \cite{Keil2016} 
where the collective oscillation of an ultracold gas may be coupled to electronic resonances
in a semiconducting micro-structure. Finally, even electrons in a two-dimensional
gas (or in a Penning trap)
may be treated with the present formalism, the two position coordinates playing
the role of the two oscillators. Magnetic fields then provide additional couplings
that also involve mixed position-momentum terms, and different dampings may be 
engineered using anisotropic textures in the direct environment of the planar trap.

The spectral analysis of entanglement and correlations between the oscillator
quadratures can be developed further. We did not explore so far the constraints
on the signal filter of Eq.\,(\ref{eq:def-linear-filter}) imposed by causality.
The hermitean correlation spectrum having optimal eigenvectors whose global phase
is arbitrary, there is some flexibility so that Kramers-Kronig relations may be not
difficult to implement.
An interesting perspective are spectral
functions that characterise the broadening of a resonance due to friction. This
concept may provide a expansion of the canonical commutator in the frequency domain
based on the correlation $\langle [x(t), p_x(t')] \rangle$. 
By linear response theory, 
this is related to the absorption spectroscopy displayed in Fig.\,\ref{fig:loss-spectra},
and would clearly provide 
a quantum reference or threshold for correlations.


\medskip
\textbf{Acknowledgements} \par 
I am indebted to Illarion Dorofeyev for instructive discussions;
his paper~\cite{Dorofeyev2013a}
triggered my interest in this problem.
Many thanks to Benjamin Schäfer, Giuseppe Cammarata, Gabriel Barton, Andreas Kurcz,
Somayyeh Nemati, 
and Janet Anders for calculations and helpful comments in various stages of this work.
We acknowledge support 
by the Deutsche Forschungsgemeinschaft through the DIP program 
(grant nos.\ Schm-1049/7-1 and Fo 703/2-1).

\appendix

\section{Derivation of the Langevin equations}

\subsection{Elimination of the bath variables}
\label{a:calculate-Langevin-eqns}

From the bath Hamiltonian~(\ref{eq:H-bath}) follows the equation
of motion for the $j$th normal mode
\begin{equation}
\dot p_j + k_j q_j = k_j c_j x
\,,\qquad
\dot q_j = p_j / m_j
\label{eq:bath-eqs}
\end{equation}
Its solution with initial conditions $(q_j(0), p_j(0))$ is
\begin{eqnarray}
q_j( t ) 
&=& q_j(0) \cos \oMega_j t 
+ \frac{ p_j(0) }{ m_j \oMega_j } \sin \oMega_j t
\nonumber\\
&& {}
+ \oMega_j c_j \!\int_0^t\!{\rm d}t' \, x(t') \sin \oMega_j ( t - t' )
\label{eq:solution-bath-mode}
\end{eqnarray}
where $\oMega_j = (k_j / m_j)^{1/2}$ is the normal mode frequency.
The derivative of this expression gives the momentum $p_j(t)$.
Since we are dealing with a linear system, both classical and quantum mechanics
give the same result. The differences originate in the initial conditions.

Inserting this result into the equation of motion for the
oscillator $(x, p_x)$, we get
\begin{equation}
\dot p_x + k_1 x =
\lambda (y-x)
- \delta k_1 x - \Gamma_1[t, x] 
+ \tilde F_1(t)
\label{eq:Langevin-0}
\end{equation}
The coordinate $y$ corresponds to the other oscillator. 
The bath-induced shift in the spring constant $k_1$ is simply
\begin{equation}
\delta k_1 = \sum_{j \in {\rm B1}} k_j c_j^2
\label{eq:}
\end{equation}
where $j \in {\rm B1}$ enumerates the set of bath modes.
The frictional force is denoted by
\begin{equation}
\Gamma_1[t, x] 
= - \int_0^t\!{\rm d}t' \, x(t') \sum_{j \in {\rm B1}} k_j \oMega_j c_j^2 \sin \oMega_j ( t - t' )
\label{eq:def-friction-Gamma1}
\end{equation}
and the Langevin force is the normal mode sum
\begin{equation}
\tilde F_1(t) = \sum_{j \in {\rm B1}} c_j \big( 
k_j q_j(0) \cos\oMega_j t
+ \oMega_j p_j(0) \sin\oMega_j t
\big)
\label{eq:def-Langevin-F1}
\end{equation}
A partial integration of Eq.\,(\ref{eq:def-friction-Gamma1}) yields the
equivalent form with a velocity-dependent friction
\begin{eqnarray}
\Gamma_1[t, x] 
&=& \int_0^t\!{\rm d}t' \, \dot x(t') \sum_{j \in {\rm B1}} k_j c_j^2 \cos \oMega_j ( t - t' )
\nonumber\\
&& {} - \sum_{j \in {\rm B1}} k_j c_j^2 \left( x(t) - x(0) \cos\oMega_j t \right)
\label{eq:}
\end{eqnarray}
In the second line, the first term with $x(t)$ cancels with the shifted spring
constant $\delta k_1$ in Eq.\,(\ref{eq:Langevin-0}). 
The second term with $x(0)$ can be combined with the
Langevin force $\tilde{F}_1(t)$, leading to the replacement
$q_j(0) \mapsto q_j(0) - x(0)$ in Eq.\,(\ref{eq:def-Langevin-F1}).
We denote in the following $F_1(t)$ the Langevin force with these
initial values.

To write the final form of the Langevin equation~(\ref{eq:Langevin-px}), 
we introduce the friction kernel
\begin{equation}
\mu_1(t - t') = \Theta(t - t') \sum_{j \in {\rm B1}} k_j c_j^2 \cos\oMega_j(t - t')
\end{equation}
A bath is by assumption a large system with a dense normal mode spectrum.
Using the mode density $\rho_1( \oMega )$ of Eq.\,(\ref{eq:def-bath-rho}),
one gets for the $\mu_1(t - t')$ the integral representation~(\ref{eq:spectrum-of-mu}) 
of the main text: it is the cosine transform of the bath spectral density 
$\rho_1( \oMega )$.

The correlation function of the Langevin force [see Eq.\,(\ref{eq:def-Langevin-F1})]
becomes with the averages of Eq.\,(\ref{eq:bath-covariances})
\begin{align}
\langle F(t), F(t') \rangle_T &= 
\sum_{j \in {\rm B1}} c_j^2 \Big\{
k_j^2 \langle (q_j(0) - x(0))^2 \rangle_T 
\cos(\oMega_j t) \cos(\oMega_j t')
\nonumber\\
& \hphantom{= \sum_{j \in {\rm B1}} c_j^2 \Big\{}
{} +
\oMega_j^2 \langle p_j(0)^2 \rangle_T 
\sin(\oMega_j t) \sin(\oMega_j t')
\Big\}
\nonumber
\\
&=
\sum_{j \in {\rm B1}} 
k_j  c_j^2 \vartheta(\oMega_j)
\cos\oMega_j (t - t')
\label{eq:}
\end{align}
Using the definition of the bath spectral density from Eq.\,(\ref{eq:def-bath-rho}),
this leads to the expression~(\ref{eq:Langevin-correlations}).
Compute also the commutator
\begin{align}
\left[ F(t), F(t') \right] &= 
\sum_{jl} c_j^2 \left\{
k_j \omega_l \left[ q_j(0) - x(0), p_l(0) \right]
\cos\omega_j t \sin\omega_l t' 
\right.
\nonumber\\
& \left. \qquad {} +
\omega_j k_l \left[ p_j(0), q_l(0) - x(0) \right]
\sin\omega_j t \cos\omega_l t' 
\right\}
\label{eq:}
\end{align}
Note that the operator $x(0)$ does not contribute. The other commutators
combine into
\begin{align}
\left[ F(t), F(t') \right] &= 
{\rm i} \hbar \sum_{j}
c_j^2 k_j \omega_j 
\sin\omega_j (t' - t)
\nonumber\\
& =
{\rm i} \hbar 
\int\limits_{0}^{\infty}\!\frac{ {\rm d}\omega }{ \pi / 2 }\,
\omega \rho( \omega ) 
\sin\omega (t' - t)
\label{eq:}
\end{align}
With the help of these expressions, one can show that the solution to the quantum
Langevin equations is such that the operators $x(t)$ and $p_x(t)$ satisfy, at each
time $t \ge 0$ the canonical commutation relations. This appears because the frictional
damping is compensated exactly by the (quantum) noise fed by the Langevin forces
and provides another application of the fluctuation--dissipation theorem in its full
quantum version.

\subsection{Ohmic friction with Drude regularisation}
\label{s:Ohm-Drude}

As a simple example, consider a friction kernel in Drude form as follows
\begin{equation}
\mu( \oMega ) = \frac{ \gamma }{ 1 - {\rm i} \oMega \tau_c}
\label{eq:Ohm-Drude-damping}
\end{equation}
It corresponds to a spectral density with a Lorentzian shape
\begin{equation}
\rho( \oMega ) = 
\frac{ \gamma / \tau_c^2 }{ \oMega^2 + 1/\tau_c^2 }
\label{eq-a:Lorentzian-bath}
\end{equation}
The memory kernel is a simple damped exponential
\begin{equation}
\mu(\tau) = \Theta(\tau) \, \frac{ \gamma }{ \tau_c }\, {\rm e}^{ - \tau / \tau_c }
\label{eq:}
\end{equation}
with a memory time $\tau_c$. The Markov limit $\tau_c \to 0$ provides a ``skew'' representation
of the $\delta(\tau)$ kernel with support in the domain $\tau \ge 0$ only.


\section{Symplectic diagonalisation}
\label{a:symplectic-diagonalisation}

The covariance matrix is the expectation value of the symmetrised correlations
$C_{ij} = \langle \hat{q}_i, \hat{q}_j \rangle_{12}$ 
where hats are used to distinguish canonical observables. 
The action of a linear canonical 
transformation $\S$ on this matrix is $\Cov = (C_{ij}) \mapsto
\S \Cov \S^{\sf T}$. We choose the lines of $\S$ to provide the coefficients 
of normal coordinates as in Eq.\,(\ref{eq:def-EPR-pair}), e.g.,
$\hat{Q} = \sum_{i} S_{1i} \hat{q}_i$. For simplicity, 
we write the coefficients (lines of $\S$) in this linear combination
just $Q^{\sf T} = (S_{1i})$, 
$P^{\sf T} = (S_{2i})$, $\ldots$ 
(without the hat). This leads to the simple
notation $\Delta \hat{Q}^2 = \langle \hat{Q}, \hat{Q} \rangle_{12} 
= Q^{\sf T} \Cov Q$.

The symplectic diagonalisation is constructed via the (non-hermitean)
eigenvalue problem
\begin{equation}
- {\rm i} \sigma \Cov v = \eta v
\label{eq:symplectic-eigen-problem}
\end{equation}
where $\sigma$ is the $4\times4$ symplectic matrix defined in 
Eq.\,(\ref{eq:def-symplectic-matrix}).
The right eigenvector $v$ encodes in its real and imaginary parts
the coefficients for the new position and momentum
coordinates: $v = Q + {\rm i} P$. 
(We assume $\eta > 0$, otherwise take the complex conjugate and work with
$v = Q - {\rm i} P$. The arbitrary phase of $v$ can be exploited to
choose $Q$ such that its overlap with the displacement coordinates
$x$ and $y$ is maximal.) 
Multiply from the left with 
$-{\rm i}(Q^{\sf T} - {\rm i} P^{\sf T})\sigma$ and get ($\sigma^2 = - \mathds{1}$)
\begin{equation}
(Q^{\sf T} - {\rm i} P^{\sf T}) \Cov (Q + {\rm i} P)
= -{\rm i} \eta (Q^{\sf T} - {\rm i} P^{\sf T}) \sigma (Q + {\rm i} P)
\label{eq:}
\end{equation}
The imaginary part of this yields
\begin{equation}
Q^{\sf T} \Cov P - P^{\sf T} \Cov Q = 
-\eta ( Q^{\sf T} \sigma Q + P^{\sf T} \sigma P )
\label{eq:}
\end{equation}
both sides are zero: on the left because $\Cov$ is symmetric, on the
right because $\sigma$ is anti-symmetric. 
Taking the real part:
\begin{equation}
Q^{\sf T} \Cov Q
+ P^{\sf T} \Cov P
= 2 \eta Q^{\sf T} \sigma P
\label{eq:sum-of-Q2-and-P2}
\end{equation}
We normalise the canonical coordinates such that 
$Q^{\sf T} \sigma P = 1$, ensuring that for the corresponding
operators, $[ \hat{Q}, \hat{P} ] = {\rm i}\hbar$. Repeating
this construction for all positive eigenvalues of 
$- {\rm i} \sigma C$, the transformation matrix is built from
the $Q$'s and $P$'s as line vectors; 
it satisfies $\S \sigma \S^{\sf T} = \sigma$.

Multiplying Eq.\,(\ref{eq:symplectic-eigen-problem}) 
with $-{\rm i}(Q^{\sf T} + {\rm i} P^{\sf T})\sigma$ from the left,
the same reasoning yields the additional information that
$Q^{\sf T} \Cov P = 0$ (the coordinates $\hat Q$, $\hat P$ are
not correlated) and that the variances 
$\Delta \hat Q^2 = Q^{\sf T} \Cov Q$ and $P^{\sf T} \Cov P$ are equal.
Therefore, from Eq.\,(\ref{eq:sum-of-Q2-and-P2})
the uncertainty product is $\Delta \hat Q \Delta \hat P = \eta$.

For the moment, we dealt with finding the normal modes of the
covariance matrix $\Cov$ itself. For the construction of an EPR pair
$\hat Q$, $\hat P$
that detects entanglement between the two oscillators, we perform
the above construction with the partially transposed covariance
matrix $\Cov^\Gamma = \Gamma \Cov \Gamma$ [Eq.\,(\ref{eq:def-PT-covariance-matrix})].
Be $v'$ thus the eigenvector of $-{\rm i}\sigma \Cov^\Gamma$ with the
smallest positive eigenvalue $\eta$. We decompose it
$Q' + {\rm i} P' = v'$ and normalise the $Q'$, $P'$ as before.
Setting then $Q = \Gamma Q'$ and $P = \Gamma P'$, we get a pair
whose variances \emph{with respect to} $\Cov$ are given by
$\Delta \hat Q = \Delta \hat P = \sqrt{\eta}$. They are not necessarily
canonically conjugate, but this is actually required to have
sub-classical correlations $\eta < \hbar / 2$.

In contrast to the above procedure, the PPT criterion performs
a positivity check of the partially transposed state \cite{Simon2000}.
This can be
done by looking for the eigenvalues of the hermitean matrix
$\Cov^\Gamma + \tfrac{\rm i}{2} \sigma$. This eigenvalue problem
\begin{align}
\left( \Cov^\Gamma + \tfrac{\rm i}{2} \sigma \right) u & = \varepsilon u
\label{eq:PPT-eigen-problem}
\\
\mbox{or } \qquad
- {\rm i} \sigma \Cov^{\Gamma} u & = - \tfrac{1}{2} u - {\rm i} \sigma \varepsilon u
\nonumber
\end{align}
typically involves complex eigenvectors $u$. They yield complex
coefficients for a linear combination of the canonical coordinates,
$\hat{a} = \sum_i u_i \hat{q}_i$,
like the familiar raising and lowering operators.

In the special case $\varepsilon = 0$, 
the $\hat{a}$ 
and $\hat{a}^\dag$ provide those lowering and raising operators 
for which $\Cov^\Gamma$ is the vacuum or ground state:
$\langle \hat{a}^\dag, \hat{a} \rangle_{12} = 0$. 
Using Eq.\,(\ref{eq:PPT-eigen-problem}),
one sees that in this case, $\Gamma u$ also 
solves~(\ref{eq:symplectic-eigen-problem})
and can be used to construct a canonical pair whose
symplectic eigenvalue is simply $\eta = 1/2$.

\medskip

%

\begin{thebibliography}{10}
\bibitem{L-z}
It is amusing to observe that this correlation is formally the same as the
  orbital angular momentum $L_z$ if the coordinates $x$, $y$ are understood as
  Cartesian components of a single particle.
\end{thebibliography}

\end{document}